\documentclass[review]{elsarticle}

\usepackage{lineno,hyperref}
\modulolinenumbers[5]

\makeatletter
\def\ps@pprintTitle{%
	\let\@oddhead\@empty
	\let\@evenhead\@empty
	\let\@oddfoot\@empty
	\let\@evenfoot\@oddfoot
}
\makeatother

\usepackage{csquotes}
\usepackage{booktabs}
\usepackage{empheq}
\usepackage[ruled,algosection]{algorithm2e}
\usepackage{geometry}%Pour ajuster les marges

\usepackage{float}%enable to use [H] for figure table,...
\usepackage{url}

\usepackage{empheq}
\usepackage{bm}
\hypersetup{
     colorlinks   = true,
     citecolor    = blue
}
\usepackage[english]{babel} %le rapport est en anglais
\addto\captionsenglish{}%Add a space after appendix

\usepackage{amsmath,amssymb}
\usepackage{cleveref}
\usepackage{epstopdf}
\usepackage{relsize}
\usepackage{stmaryrd} % For \llbracket and \rrbracket
\usepackage{subcaption}
\usepackage{multirow}
\usepackage{array,makecell}
\usepackage{import}

\usepackage{pifont}% http://ctan.org/pkg/pifont

\usepackage{amsthm}

\newdefinition{rmk}{Remark}[section]
\newproof{pf}{Proof}
\newproof{pot}{Proof of Theorem \ref{thm2}}
\newdefinition{definition}{Definition}[section]

\usepackage[binary-units]{siunitx}

\usepackage{enumitem}

\usepackage{scalerel}

\usepackage{amsthm}

\usepackage{tikz}
\usetikzlibrary{intersections,arrows.meta,calc,patterns,decorations.pathmorphing,decorations.markings,shapes,fit}
\usetikzlibrary{plotmarks}
\usepackage{xparse}
\DeclareDocumentCommand\Courbe{ m O{} }{(%
	\tikz[baseline=-\the\dimexpr\fontdimen22\textfont2\relax,inner sep=0pt] {%
		\draw[line width=1pt,text height = \textheight,#1] plot coordinates {(0,0)} -- plot[mark options={solid}, mark size=2pt,#2] coordinates {(2mm,0)} -- plot coordinates {(4mm,0)};%
	})%
}%
\DeclareRobustCommand{\Courbearrow}[1]{[\tikz[baseline=-\the\dimexpr\fontdimen22\textfont2\relax,inner sep=0pt,>=stealth] \draw[->,line width=1pt,#1](0,0) -- (3mm,0);]}
\DeclareRobustCommand{\Courbearrow}[1]{[\tikz[baseline=-\the\dimexpr\fontdimen22\textfont2\relax,inner sep=0pt,>=stealth] \draw[->,line width=1pt,#1](0,0) -- (3mm,0);]}

\usepackage{xcolor}
\colorlet{NLddl1}{black!50!red}
\colorlet{NLddl2}{blue}%{0.62500,0.10000,0.15000}%
\colorlet{NLddl3}{black!30!green}%{0.75000,0.20000,0.30000}%
\colorlet{NLddl4}{red!100!blue!60!}%{0.87500,0.30000,0.45000}%
\colorlet{NLddl5}{red!35!green!30!blue!50!}%{0.45,0.76,0.98}%

\colorlet{mainF25}{black!50!red}
\colorlet{BifF25}{black!50!green}
\colorlet{mainF15}{blue}%{0.62500,0.10000,0.15000}%
\colorlet{BifF15}{red!50}
\colorlet{mainF5}{black!50}%{0.87500,0.30000,0.45000}%
\colorlet{BifF5}{red!50!blue!50}%{0.43823,0.61177,0.86615}%
\colorlet{360HBM}{black!50!green!}
\colorlet{decouple}{white!40!red}
\definecolor{lightred}{RGB}{0,110,0}%%Vert foncé
\definecolor{darkred}{RGB}{0,0,0}

\colorlet{Coulomb}{white!40!red}

\usepackage{algorithmic}

\usepackage{mathptmx}

\title{A Virtual Acoustic Black Hole on a Cantilever Beam}
\usepackage{xcolor}
\definecolor{LTDSdarkgreen}{RGB}{0,130,0}%Vert foncé
\definecolor{Magenta}{RGB}{130,0,130}%Vert foncé
\usepackage{graphicx}
\begin{document}
\begin{frontmatter}
% Affiliations

\author[1]{Samuel Quaegebeur\corref{mycorrespondingauthor}}
\cortext[mycorrespondingauthor]{Corresponding author}
\ead{squaegebeur@uliege.be}
\author[1]{Ghislain Raze}
\author[2]{Li Cheng}
\author[1]{Gaëtan Kerschen}
\address[1]{Department of Aerospace and Mechanical Engineering,
	University of Liège, Allée de la Découverte 9
	B-4000 Liège, Belgium}
\address[2]{Department of Mechanical Engineering,
	The Hong Kong Polytechnic University, Hung Hom
	Kowloon, Hong Kong SAR, P. R. China}

\begin{abstract}
An acoustic black hole (ABH) consists of a tapered structure whose thickness follows a power-law profile. When attached to a host structure, an ABH localizes and traps the vibrational energy, which can then be dissipated through, e.g., a damping layer. However, effective vibration mitigation is known to occur only above a cut-on frequency which is inversely proportional to the length of the tapered structure. In this context, the main thrust of this paper is to replace a mechanical ABH by a digital controller so as to create a so-called \textit{virtual acoustic black hole} (VABH), thus, freeing the ABH from possible mechanical constraints (e.g., compactness, manufacturing and fatigue issues). The proposed VABH is first detailed theoretically. The salient features and performance of the VABH are then demonstrated both numerically and experimentally using a cantilever beam as a host structure. Eventually, it is shown that the VABH significantly enlarges the applicability of the concept of an ABH.

\end{abstract}


\end{frontmatter}

\linenumbers
\section{Introduction}
\begin{figure}[H]
    \centering
    \begin{subfigure}{\textwidth}
    \centering
    \includegraphics[scale=0.1]{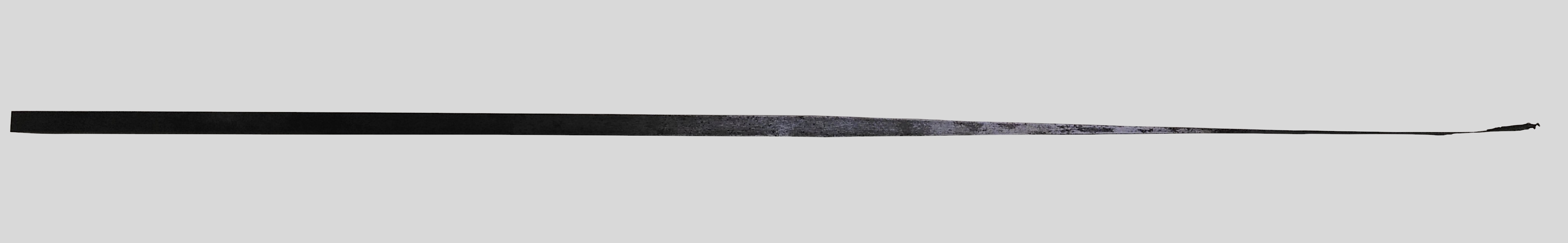}
    \caption{Cantilever beam equipped with an ABH.}
    \label{subfig:length meca ABH}
    \end{subfigure}

\par\bigskip % force a bit of vertical whitespace
%We now put the shape of the sector

    \begin{subfigure}{0.5\textwidth}
    \centering
    \includegraphics[height=5.5cm]{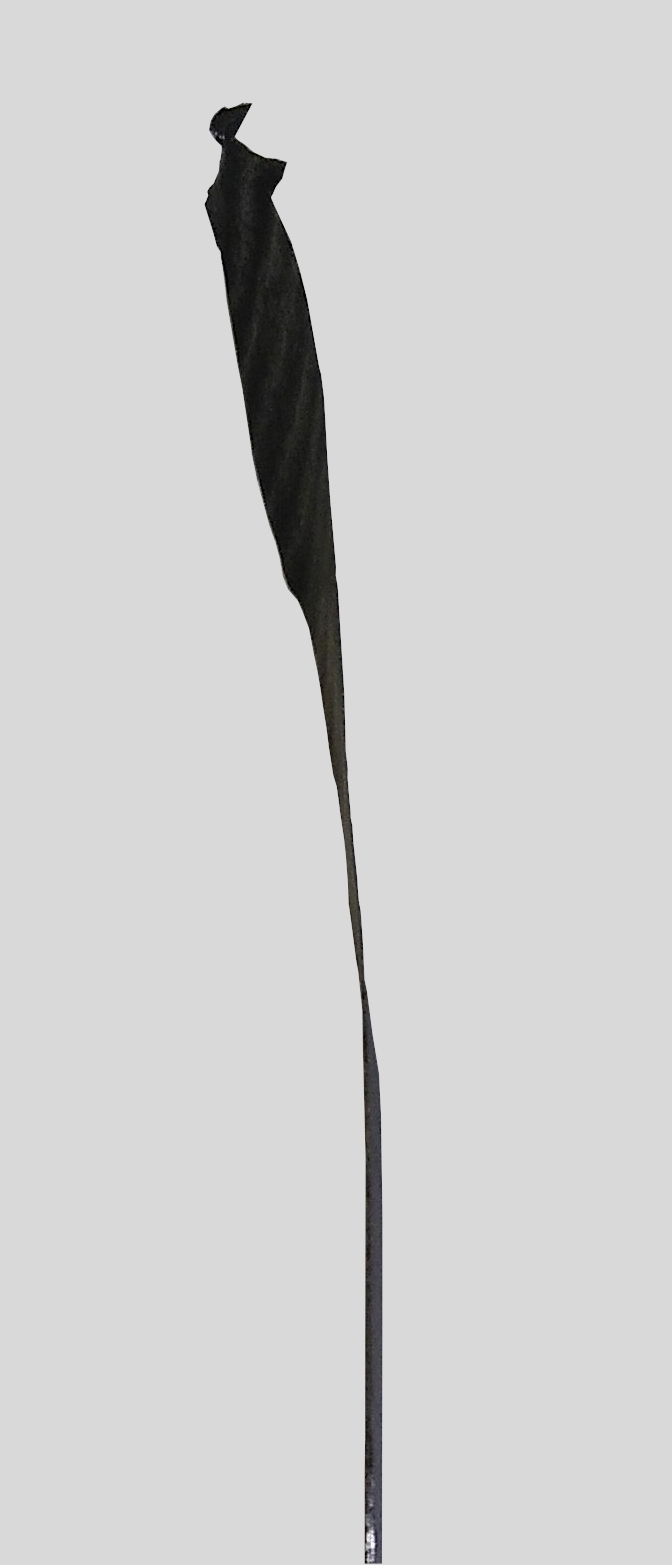}
    \caption{Close-up at the tip, side view.}
    \label{subfig:side2 ABH}
    \end{subfigure}%
        \begin{subfigure}{0.5\textwidth}
    \centering
    \includegraphics[height=5.5cm]{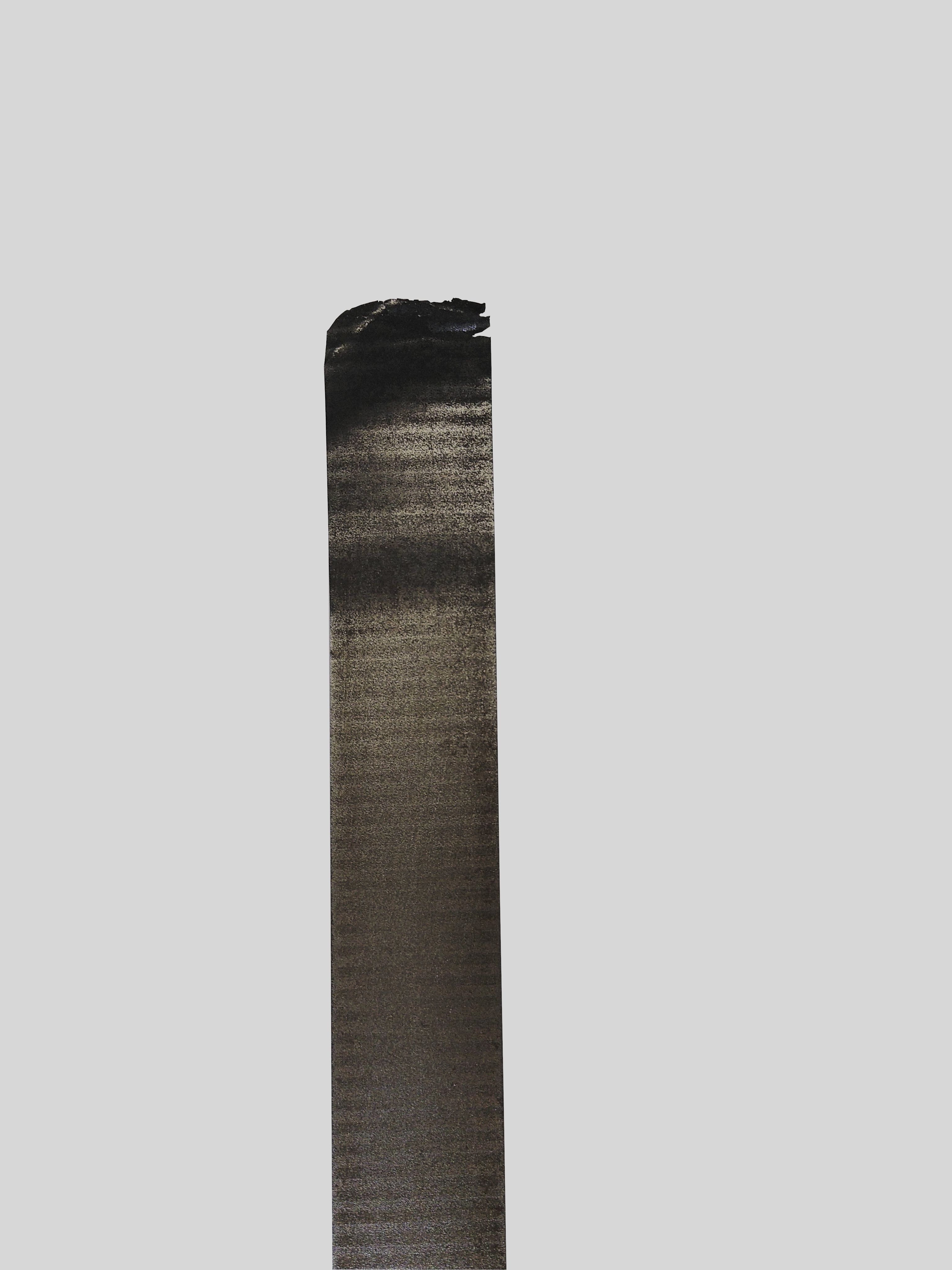}
    \caption{Close-up at the tip, front view. }
    \label{subfig:side 1 ABH}
    \end{subfigure}
    \caption{Beam with a $\SI{35}{\centi\meter}$-long ABH. Due to the small thickness, the tip was crooked during experimental tests.}
    \label{fig:ABH meca}
\end{figure}

The acoustic black hole (ABH) effect was first proposed by Mironov in~\cite{mironov_propagation_1988}. The mechanical device consists of a tapered wedge beam with a variable thickness following a power-law function. Because the group velocity of a flexural wave is proportional to the square root of the thickness, an acoustic trap with - in principle - no reflection can be achieved through a mechanical device whose tip has a zero thickness. In view of unavoidable manufacturing tolerances, the achievable absorption performance necessarily decreases. To enhance the ABH effect, different strategies were proposed during the last two decades, namely the addition of a damping layer~\cite{krylov_acoustic_2004,denis_modal_2014} or the use of an extended platform at the end of the tapered beam~\cite{tang_enhanced_2017}. The different features of an ABH along with the existing body of literature are explained in detail in the review paper~\cite{pelat_acoustic_2020}. One of the most important parameters is the cut-on frequency above which the ABH starts to be effective. Because this frequency is inversely proportional to the length of the tapered wedge beam, a particularly long and thin ABH is necessary for vibration mitigation at low excitation frequencies (say below 100 Hz). Several hindrances may arise from this feature, e.g., the manufacturing process can be challenging, and the mechanical system can become cumbersome and brittle, as shown in Figure \ref{fig:ABH meca}. Attempts to obtain a more compact system \cite{lee_vibration_2017,zhou_resonant_2018} or to increase the fatigue limits~\cite{fatigue_ABH_keys} were carried out. Different studies were also conducted to achieve vibration reduction below the cut-on frequency of the system. For instance, references~\cite{li_experimental_2021,li_combining_2021} exploit nonlinearities to transfer energy from low to high frequencies where the ABH is known to be effective. Other studies combine the ABH effect with either piezoelectric transducers connected to linear/nonlinear shunt circuits ~\cite{zhang_electromechanical_2020,li_energy_2022,zhang_nonlinear_2022} or feedforward/feedback control laws~\cite{hook_control_2022,cheer_active_2021,hook_active_2022}. However, the manufacturing process and the fragility of the ABH still remain important bottlenecks.  

%3 possibilités
%\begin{figure}[H]
    %\centering
   % \includegraphics[scale=0.1]{body/images/Mechanical_ABH/ABH_12.png}
  %  \caption{Beam with an ABH. The tapered wedge is close to $\SI{50}{\centi\meter}$ long. Due to the very thin thickness, the tip got crooked during experimental test. }
 %   \label{fig:image ABH meca}
%\end{figure}

%\begin{figure}[H]
 %   \centering
  %  \includegraphics[scale=0.1]{body/images/Mechanical_ABH/ABH_blanc.png}
   % \caption{Beam with an ABH. The tapered wedge is close to $\SI{50}{\centi\meter}$ long. Due to the very thin thickness, the tip got crooked during experimental test. }
%    \label{fig:image ABH meca}
%\end{figure}

\begin{figure}[H]
\centering
\includegraphics[scale=0.8,trim={2cm} {0} {0} {0},clip]{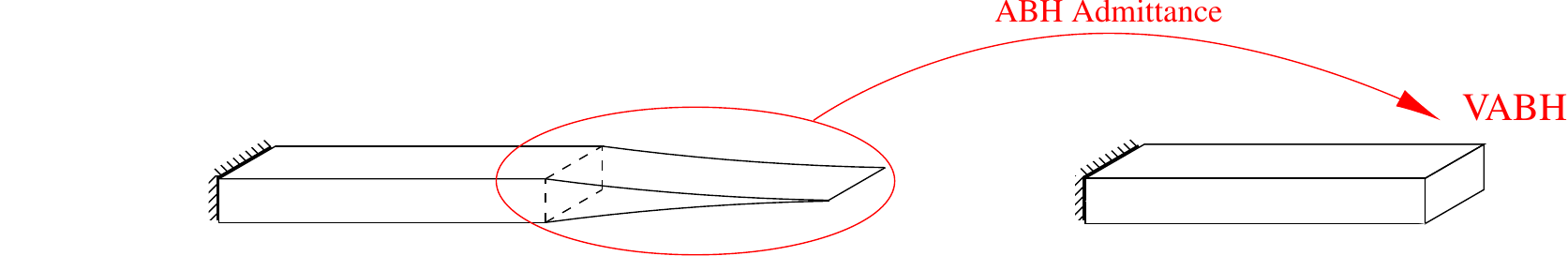}
\caption{Schematics of the virtual acoustic black hole. }
\label{fig:intro VABH}
\end{figure}

The present study finds its roots in the recent development of digital vibration absorbers \cite{fleming_synthetic_2000,matten_synthetic_2014,sugino_design_2018,yi_programmable_2020} or energy harvesters~\cite{sugino_analysis_2018}. Because a digital controller can synthesize virtually any impendance function (even nonlinear ones ~\cite{raze_digital_2019}), it can advantageously replace analog shunt circuits for piezoelectric vibration absorption~\cite{forward_electronic_1979,hagood_damping_1991}. In this paper, we propose the novel concept of a \textit{virtual ABH} (VABH) where the ABH is replaced by a digital controller, see Figure~\ref{fig:intro VABH}. Employing this strategy paves the way for the practical realization of very long and thin ABHs without suffering from the aforementioned mechanical constraints. An interesting, though fundamentally different, study suggesting the use of programmable piezoelectric shunt circuits for mimicking the dynamical characteristics of a mechanical ABH is that of Sugino and co-workers~\cite{sugino_spatially_2022}. In their work, the impedance applied to each unit cell is varied so as to control the local dispersion properties of the structure and reproduce the wavelength compression of an ABH.

The paper is organized as follows. Section~\ref{sec:formulation ABH} establishes the state space formulation of the ABH effect that will serve as a basis for the feedback control law. The VABH and its salient dynamical features are studied numerically in  Section~\ref{sec:numerical features}.  Section~\ref{sec:final app} discusses the experimental setup and the requirements associated with the VABH. This Section also carries out a correlation with numerical simulations. Section~\ref{sec:performances ABH} presents the performance of an experimental VABH. Thanks to its virtual nature, ABH with different properties including different materials and lengths can easily be compared. Finally, the conclusions of this study are drawn in Section 6.

\section{Formulation of the ABH dynamics as a feedback function}\label{sec:formulation ABH}
\subsection{Finite element modeling}
\begin{figure}[H]
	\centering
	\includegraphics[scale=1]{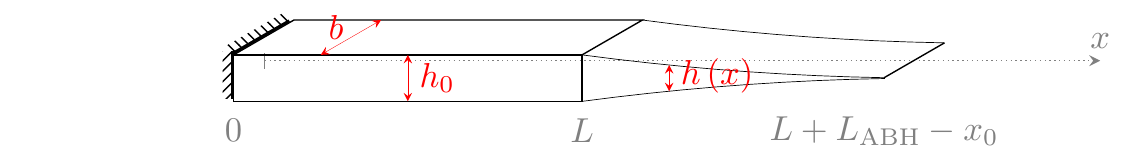}
	\caption{Schematics of the cantilever beam with a tapered wedge profile.}
	\label{fig:schemabeam+ABH}
\end{figure}

The purpose of this Section is to formulate the ABH effect as a mechanical feedback function. Figure~\ref{fig:schemabeam+ABH} depicts a tapered wedge beam attached to a uniform cantilever beam. The uniform beam is characterized by its material and geometry; $h_{0}$, $L$ and $b$ denote the thickness, length and width, respectively.  The tapered wedge beam has a length $L_{\mathrm{ABH}}-x_{0}$, where $x_{0}$ corresponds to the tapered wedge beam's truncation. Its width is equal to $b$; its thickness follows the law
\begin{equation}
	h\left(x\right)=h_{0}\left(\dfrac{L+L_{\mathrm{ABH}}-x}{L_{\mathrm{ABH}}}\right)^{m},\,x\in\left[L,L+L_{\mathrm{ABH}}-x_{0}\right],
\end{equation}
where $m$ is an integer greater than 1. Due to the truncation $x_{0}$, the tip of the tapered wedge beam has a residual thickness equal to $h_{0}\left(\frac{x_{0}}{L_{\mathrm{ABH}}}\right)^{m}$. 

The finite element method is employed to model the coupled system. In what follows, the tapered wedge and cantilever beams are identified with superscripts $\mathrm{tb}$ and $\mathrm{b}$, respectively. For both systems, the internal and boundary nodes are denoted by the subscripts $\mathrm{I}$ and $\mathrm{B}$, respectively. The equations of motion read
\begin{subequations}
	\begin{equation}\label{eq:dynamics beam}
		\begin{bmatrix}
			\mathbf{M}^{\mathrm{b}}_{\mathrm{II}}&\mathbf{M}^{\mathrm{b}}_{\mathrm{IB}}\\
			\mathbf{M}^{\mathrm{b}}_{\mathrm{BI}}&\mathbf{M}^{\mathrm{b}}_{\mathrm{BB}}
		\end{bmatrix}
		\begin{bmatrix}
			\ddot{\mathbf{x}}^{\mathrm{b}}_{\mathrm{I}}\\
			\ddot{\mathbf{x}}_{\mathrm{B}}
		\end{bmatrix}+
		\begin{bmatrix}
			\mathbf{C}^{\mathrm{b}}_{\mathrm{II}}&\mathbf{C}^{\mathrm{b}}_{\mathrm{IB}}\\
			\mathbf{C}^{\mathrm{b}}_{\mathrm{BI}}&\mathbf{C}^{\mathrm{b}}_{\mathrm{BB}}
		\end{bmatrix}
		\begin{bmatrix}
			\dot{\mathbf{x}}^{\mathrm{b}}_{\mathrm{I}}\\
			\dot{\mathbf{x}}_{\mathrm{B}}
		\end{bmatrix}+
		\begin{bmatrix}
			\mathbf{K}^{\mathrm{b}}_{\mathrm{II}}&\mathbf{K}^{\mathrm{b}}_{\mathrm{IB}}\\
			\mathbf{K}^{\mathrm{b}}_{\mathrm{BI}}&\mathbf{K}^{\mathrm{b}}_{\mathrm{BB}}
		\end{bmatrix}
		\begin{bmatrix}
			\mathbf{x}^{\mathrm{b}}_{\mathrm{I}}\\
			\mathbf{x}_{\mathrm{B}}
		\end{bmatrix}
		=\begin{bmatrix}
			\mathbf{f}_{\mathrm{ext,I}} \\
			\mathbf{f}_{\mathrm{ext,B}}+\mathbf{f}_{\mathrm{tb}\rightarrow\mathrm{b}}
		\end{bmatrix}
	\end{equation}
	\begin{equation}\label{eq:ABH equation motion}
		\begin{bmatrix}
			\mathbf{M}^{\mathrm{tb}}_{\mathrm{II}}&\mathbf{M}^{\mathrm{tb}}_{\mathrm{I B}}\\
			\mathbf{M}^{\mathrm{tb}}_{\mathrm{BI}}&\mathbf{M}^{\mathrm{tb}}_{\mathrm{BB}}
		\end{bmatrix}
		\begin{bmatrix}
			\ddot{\mathbf{x}}^{\mathrm{tb}}_{\mathrm{I}}\\
			\ddot{\mathbf{x}}_{\mathrm{B}}
		\end{bmatrix}+
		\begin{bmatrix}
			\mathbf{C}^{\mathrm{tb}}_{\mathrm{II}}&\mathbf{C}^{\mathrm{tb}}_{\mathrm{I B}}\\
			\mathbf{C}^{\mathrm{tb}}_{\mathrm{B I}}&\mathbf{C}^{\mathrm{tb}}_{\mathrm{BB}}
		\end{bmatrix}
		\begin{bmatrix}
			\dot{\mathbf{x}}^{\mathrm{tb}}_{\mathrm{I}}\\
			\dot{\mathbf{x}}_{\mathrm{B}}
		\end{bmatrix}+
		\begin{bmatrix}
			\mathbf{K}^{\mathrm{tb}}_{\mathrm{II}}&\mathbf{K}^{\mathrm{tb}}_{\mathrm{I B}}\\
			\mathbf{K}^{\mathrm{tb}}_{\mathrm{BI}}&\mathbf{K}^{\mathrm{tb}}_{\mathrm{BB}}
		\end{bmatrix}
		\begin{bmatrix}
			\mathbf{x}^{\mathrm{tb}}_{\mathrm{I}}\\
			\mathbf{x}_{\mathrm{B}}
		\end{bmatrix}
		=\begin{bmatrix}
			\mathbf{0} \\
			-\mathbf{f}_{\mathrm{tb}\rightarrow\mathrm{b}}
		\end{bmatrix},
	\end{equation}
\end{subequations}
where $\mathbf{M}$, $\mathbf{C}$, and $\mathbf{K}$ represent the mass, damping and stiffness matrices, respectively. The vector $\mathbf{f}_{\mathrm{ext}}$ represents the external forces applied to the cantilever beam. The tapered wedge beam applies forces and torques, $\mathbf{f}_{\mathrm{tb}\rightarrow\mathrm{b}}$, to the cantilever beam. Based on Newton's third law, opposite forces and torques are applied to the tapered wedge beam. If we were to consider the equations of motion of the assembled system, these terms would cancel out. However, our objective is to write the ABH effect generated by the tapered wedge beam as a feedback transfer function. The vector $\mathbf{f}_{\mathrm{tb}\rightarrow\mathrm{b}}$ describes this effect; it thus plays a key role in our analysis.

\subsection{State-space formulation}

\begin{figure}[H]
	\centering
	\includegraphics[scale=1]{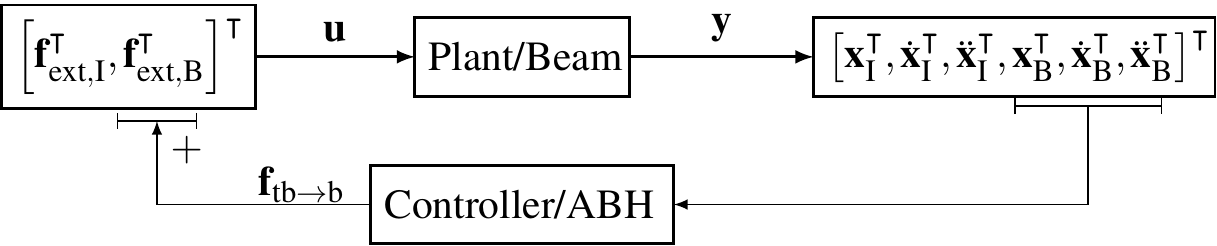}
	\caption{Representation of beam/ABH system with a feedback diagram.}
	\label{fig:state space formulation}
\end{figure}

According to control theory, the cantilever beam and ABH are referred to as "plant" and "controller", respectively, as in Figure~\ref{fig:state space formulation}. The equations of motion of the cantilever beam are recast into state-space formulation:
\begin{equation}\label{eq:state space formulation}
	\left\lbrace
	\begin{array}{lll}
		\dot{\mathbf{x}}=\mathbf{E}^{\mathrm{b}}\mathbf{x}+\mathbf{F}^{\mathrm{b}}\mathbf{u} \\
		\mathbf{y} = \mathbf{G}^{\mathrm{b}}\mathbf{x}+\mathbf{H}^{\mathrm{b}}\mathbf{u},
	\end{array}
	\right.
\end{equation}
The vectors $\mathbf{u}=\left[\mathbf{f}_{\mathrm{ext,I}}^{\intercal},\mathbf{f}_{\mathrm{ext,B}}^{\intercal}\right]^{\intercal}$, $\mathbf{y}=\left[\mathbf{x}_{\mathrm{I}}^{\intercal},\dot{\mathbf{x}}_{\mathrm{I}}^{\intercal},\ddot{\mathbf{x}}_{\mathrm{I}}^{\intercal},\mathbf{x}_{\mathrm{B}}^{\intercal},\dot{\mathbf{x}}_{\mathrm{B}}^{\intercal},\ddot{\mathbf{x}}_{\mathrm{B}}^{\intercal}\right]^{\intercal}$, and $\mathbf{x}$ represent the input, output and state-space variables, respectively. The state-space matrices are
\begin{subequations}
	\begin{equation}
		\mathbf{E}^{\mathrm{b}}=\begin{bmatrix}
			\mathbf{0}_{N_{\mathrm{b}}\mathrm{x}N_{\mathrm{b}}}&\mathbf{I}_{N_{\mathrm{b}}\mathrm{x}N_{\mathrm{b}}} \\
			-\left(\mathbf{M}^{\mathrm{b}}\right)^{-1}\mathbf{K}^{\mathrm{b}}&-\left(\mathbf{M}^{\mathrm{b}}\right)^{-1}\mathbf{C}^{\mathrm{b}}
		\end{bmatrix}
,		\mathbf{F}^{\mathrm{b}}=\begin{bmatrix}
			\mathbf{0}_{N_{\mathrm{b}}\mathrm{x}N_{\mathrm{b}}} \\
			\left(\mathbf{M}^{\mathrm{b}}\right)^{-1}
		\end{bmatrix}
	\end{equation}
	\begin{equation}
		\mathbf{G}^{\mathrm{b}}=\begin{bmatrix}
			\mathbf{I}_{N_{\mathrm{b}}\mathrm{x}N_{\mathrm{b}}}&\mathbf{0}_{N_{\mathrm{b}}\mathrm{x}N_{\mathrm{b}}} \\
			\mathbf{0}_{N_{\mathrm{b}}\mathrm{x}N_{\mathrm{b}}}&\mathbf{I}_{N_{\mathrm{b}}\mathrm{x}N_{\mathrm{b}}} \\
			-\left(\mathbf{M}^{\mathrm{b}}\right)^{-1}\mathbf{K}^{\mathrm{b}} &-\left(\mathbf{M}^{\mathrm{b}}\right)^{-1}\mathbf{C}^{\mathrm{b}}
		\end{bmatrix}
,
		\mathbf{H}^{\mathrm{b}}=\begin{bmatrix}
			\mathbf{0}_{N_{\mathrm{b}}\mathrm{x}N_{\mathrm{b}}} \\
			\mathbf{0}_{N_{\mathrm{b}}\mathrm{x}N_{\mathrm{b}}} \\
			\left(\mathbf{M}^{\mathrm{b}}\right)^{-1}
		\end{bmatrix}
	\end{equation}
\end{subequations}
where $N_{\mathrm{b}}$ denotes the number of degrees of freedom of the cantilever beam.  

To use the tapered wedge beam as a feedback function, its input vector should contain the displacement $\mathbf{x}_{\mathrm{B}}$, velocity $\dot{\mathbf{x}}_{\mathrm{B}}$ and acceleration $\ddot{\mathbf{x}}_{\mathrm{B}}$ at the boundary of the beam, and its output should correspond to the forces and torques at the boundary, $\mathbf{f}_{\mathrm{tb}\rightarrow\mathrm{b}}$. The equations of motion of the tapered wedge beam are 
\begin{equation}\label{eq:SS ABH}
	\left\lbrace
	\begin{array}{lll}
		\dfrac{\mathrm{d}}{\mathrm{d}t}
		\begin{bmatrix}
			\mathbf{x}^{\mathrm{tb}}_{\mathrm{I}}\\
			\dot{\mathbf{x}}^{\mathrm{tb}}_{\mathrm{I}}
		\end{bmatrix}=\mathbf{E}^{\mathrm{tb}}\begin{bmatrix}
			\mathbf{x}^{\mathrm{tb}}_{\mathrm{I}}\\
			\dot{\mathbf{x}}^{\mathrm{tb}}_{\mathrm{I}}
		\end{bmatrix}+\mathbf{F}^{\mathrm{tb}}\begin{bmatrix}
			\mathbf{x}_{\mathrm{B}}\\
			\dot{\mathbf{x}}_{\mathrm{B}} \\
			\ddot{\mathbf{x}}_{\mathrm{B}}
		\end{bmatrix} \\
		\mathbf{f}_{\mathrm{tb}\rightarrow\mathrm{b}} = \mathbf{G}^{\mathrm{tb}}\begin{bmatrix}
			\mathbf{x}^{\mathrm{tb}}_{\mathrm{I}}\\
			\dot{\mathbf{x}}^{\mathrm{tb}}_{\mathrm{I}}
		\end{bmatrix}+\mathbf{H}^{\mathrm{tb}}\begin{bmatrix}
			\mathbf{x}_{\mathrm{B}}\\
			\dot{\mathbf{x}}_{\mathrm{B}} \\
			\ddot{\mathbf{x}}_{\mathrm{B}}
		\end{bmatrix}
	\end{array}
	\right.
\end{equation}
with
\begin{subequations}
	\begin{equation}
		\mathbf{E}^{\mathrm{tb}}=\begin{bmatrix}
			\mathbf{0}&\mathbf{I} \\
			-\left(\mathbf{M}^{\mathrm{tb}}_{\mathrm{II}}\right)^{-1}\mathbf{K}^{\mathrm{tb}}_{\mathrm{II}}&-\left(\mathbf{M}^{\mathrm{tb}}_{\mathrm{II}}\right)^{-1}\mathbf{C}^{\mathrm{tb}}_{\mathrm{II}}
		\end{bmatrix}
,
		\mathbf{F}^{\mathrm{tb}}=\begin{bmatrix}
			\mathbf{0}&\mathbf{0}&\mathbf{0} \\
			-\left(\mathbf{M}^{\mathrm{tb}}_{\mathrm{II}}\right)^{-1}\mathbf{K}^{\mathrm{tb}}_{\mathrm{IB}}&-\left(\mathbf{M}^{\mathrm{tb}}_{\mathrm{II}}\right)^{-1}\mathbf{C}^{\mathrm{tb}}_{\mathrm{IB}}&-\left(\mathbf{M}^{\mathrm{tb}}_{\mathrm{II}}\right)^{-1}\mathbf{M}^{\mathrm{tb}}_{\mathrm{IB}}
		\end{bmatrix}
	\end{equation}
	\begin{equation}
		\mathbf{G}^{\mathrm{tb}}=\begin{bmatrix}
			\mathbf{K}^{\mathrm{tb}}_{\mathrm{BI}}-\mathbf{M}^{\mathrm{tb}}_{\mathrm{BI}}\left(\mathbf{M}^{\mathrm{tb}}_{\mathrm{II}}\right)^{-1}\mathbf{K}^{\mathrm{tb}}_{\mathrm{II}} &\mathbf{C}^{\mathrm{tb}}_{\mathrm{BI}}-\mathbf{M}^{\mathrm{tb}}_{\mathrm{BI}}\left(\mathbf{M}^{\mathrm{tb}}_{\mathrm{II}}\right)^{-1}\mathbf{C}^{\mathrm{tb}}_{\mathrm{II}}
		\end{bmatrix}
	\end{equation}
	\begin{equation}
		\mathbf{H}^{\mathrm{tb}}=\begin{bmatrix}
			\mathbf{K}^{\mathrm{tb}}_{\mathrm{BB}}-\mathbf{M}^{\mathrm{tb}}_{\mathrm{BI}}\left(\mathbf{M}^{\mathrm{tb}}_{\mathrm{II}}\right)^{-1}\mathbf{K}^{\mathrm{tb}}_{\mathrm{IB}} &\mathbf{C}^{\mathrm{tb}}_{\mathrm{BB}}-\mathbf{M}^{\mathrm{tb}}_{\mathrm{BI}}\left(\mathbf{M}^{\mathrm{tb}}_{\mathrm{II}}\right)^{-1}\mathbf{C}^{\mathrm{tb}}_{\mathrm{IB}} &\mathbf{M}^{\mathrm{tb}}_{\mathrm{BB}}-\mathbf{M}^{\mathrm{tb}}_{\mathrm{BI}}\left(\mathbf{M}^{\mathrm{tb}}_{\mathrm{II}}\right)^{-1}\mathbf{M}^{\mathrm{tb}}_{\mathrm{IB}}
		\end{bmatrix}.
	\end{equation}
\end{subequations}

In practice, only one type of output is measured from the plant (either displacement, velocity or acceleration). If the output acceleration is measured (as is the case in this study), the plant velocity and displacement can be obtained through single and double integration, respectively. The state-space model can thus be built as
\begin{equation}\label{eq:SS ABH2}
	\left\lbrace
	\begin{array}{lll}
		\dfrac{\mathrm{d}}{\mathrm{d}t}\begin{bmatrix}
			\mathbf{x}^{\mathrm{tb}}_{\mathrm{I}}\\
			\mathbf{x}_{\mathrm{B}} \\
			\dot{\mathbf{x}}^{\mathrm{tb}}_{\mathrm{I}} \\
			\dot{\mathbf{x}}_{\mathrm{B}}
		\end{bmatrix}=\mathbf{E}^{\mathrm{tb}}\begin{bmatrix}
			\mathbf{x}^{\mathrm{tb}}_{\mathrm{I}}\\
			\mathbf{x}_{\mathrm{B}}\\
			\dot{\mathbf{x}}^{\mathrm{tb}}_{\mathrm{I}}\\
			\dot{\mathbf{x}}_{\mathrm{B}}
		\end{bmatrix}+\mathbf{F}^{\mathrm{tb}}
		\ddot{\mathbf{x}}_{\mathrm{B}}, \\
		\mathbf{f}_{\mathrm{tb}\rightarrow\mathrm{b}} = \mathbf{G}^{\mathrm{tb}}\begin{bmatrix}
			\mathbf{x}^{\mathrm{tb}}_{\mathrm{I}}\\
			\mathbf{x}_{\mathrm{B}}\\
			\dot{\mathbf{x}}^{\mathrm{tb}}_{\mathrm{I}}\\
			\dot{\mathbf{x}}_{\mathrm{B}}
		\end{bmatrix}+\mathbf{H}^{\mathrm{tb}}
		\ddot{\mathbf{x}}_{\mathrm{B}},
	\end{array}
	\right.
\end{equation}
with 
\begin{subequations}
	\begin{equation}
		\mathbf{E}^{\mathrm{tb}}=\begin{bmatrix}
			\mathbf{0}&\mathbf{0}&\mathbf{I}&\mathbf{0} \\
			\mathbf{0}&\mathbf{0}&\mathbf{0}&\mathbf{I} \\
			-\left(\mathbf{M}^{\mathrm{tb}}_{\mathrm{II}}\right)^{-1}\mathbf{K}^{\mathrm{tb}}_{\mathrm{II}} & -\left(\mathbf{M}^{\mathrm{tb}}_{\mathrm{II}}\right)^{-1}\mathbf{K}^{\mathrm{tb}}_{\mathrm{IB}} &-\left(\mathbf{M}^{\mathrm{tb}}_{\mathrm{II}}\right)^{-1}\mathbf{C}^{\mathrm{tb}}_{\mathrm{II}} & -\left(\mathbf{M}^{\mathrm{tb}}_{\mathrm{II}}\right)^{-1}\mathbf{C}^{\mathrm{tb}}_{\mathrm{IB}} \\
			\mathbf{0}&\mathbf{0}&\mathbf{0}&\mathbf{0}
		\end{bmatrix}
,
		\mathbf{F}^{\mathrm{tb}}=\begin{bmatrix}
			\mathbf{0} \\
			\mathbf{0} \\
			-\left(\mathbf{M}^{\mathrm{tb}}_{\mathrm{II}}\right)^{-1}\mathbf{M}^{\mathrm{tb}}_{\mathrm{IB}} \\
			\mathbf{I}
		\end{bmatrix}
	\end{equation}
	% \begin{equation}
		% 	\mathbf{G}=\begin{bmatrix}
			% 		\mathbf{K}^{\mathrm{tb}}_{\mathrm{BI}}-\mathbf{M}^{\mathrm{tb}}_{\mathrm{BI}}\left(\mathbf{M}^{\mathrm{tb}}_{\mathrm{II}}\right)^{-1}\mathbf{K}^{\mathrm{tb}}_{\mathrm{II}} 
			%            &\mathbf{K}^{\mathrm{tb}}_{\mathrm{BB}}-\mathbf{M}^{\mathrm{tb}}_{\mathrm{BI}}\left(\mathbf{M}^{\mathrm{tb}}_{\mathrm{II}}\right)^{-1}\mathbf{K}^{\mathrm{tb}}_{\mathrm{IB}}
			%            &\mathbf{C}^{\mathrm{tb}}_{\mathrm{BI}}-\mathbf{M}^{\mathrm{tb}}_{\mathrm{BI}}\left(\mathbf{M}^{\mathrm{tb}}_{\mathrm{II}}\right)^{-1}\mathbf{C}^{\mathrm{tb}}_{\mathrm{II}}
			%            & \mathbf{C}^{\mathrm{tb}}_{\mathrm{BB}}-\mathbf{M}^{\mathrm{tb}}_{\mathrm{BI}}\left(\mathbf{M}^{\mathrm{tb}}_{\mathrm{II}}\right)^{-1}\mathbf{C}^{\mathrm{tb}}_{\mathrm{IB}}
			% 	\end{bmatrix}
		% \end{equation}
	\begin{multline}
		\mathbf{G}^{\mathrm{tb}}=\left[
		\begin{array}{cc} \mathbf{K}^{\mathrm{tb}}_{\mathrm{BI}}-\mathbf{M}^{\mathrm{tb}}_{\mathrm{BI}}\left(\mathbf{M}^{\mathrm{tb}}_{\mathrm{II}}\right)^{-1}\mathbf{K}^{\mathrm{tb}}_{\mathrm{II}} 
			&\mathbf{K}^{\mathrm{tb}}_{\mathrm{BB}}-\mathbf{M}^{\mathrm{tb}}_{\mathrm{BI}}\left(\mathbf{M}^{\mathrm{tb}}_{\mathrm{II}}\right)^{-1}\mathbf{K}^{\mathrm{tb}}_{\mathrm{IB}}\end{array}\right.
		\\
		\left.
		\begin{array}{cc} \mathbf{C}^{\mathrm{tb}}_{\mathrm{BI}}-\mathbf{M}^{\mathrm{tb}}_{\mathrm{BI}}\left(\mathbf{M}^{\mathrm{tb}}_{\mathrm{II}}\right)^{-1}\mathbf{C}^{\mathrm{tb}}_{\mathrm{II}}
			& \mathbf{C}^{\mathrm{tb}}_{\mathrm{BB}}-\mathbf{M}^{\mathrm{tb}}_{\mathrm{BI}}\left(\mathbf{M}^{\mathrm{tb}}_{\mathrm{II}}\right)^{-1}\mathbf{C}^{\mathrm{tb}}_{\mathrm{IB}}\end{array}
		\right]
	\end{multline}
	\begin{equation}
		\mathbf{H}^{\mathrm{tb}}=
		\mathbf{M}^{\mathrm{tb}}_{\mathrm{BB}}-\mathbf{M}^{\mathrm{tb}}_{\mathrm{BI}}\left(\mathbf{M}^{\mathrm{tb}}_{\mathrm{II}}\right)^{-1}\mathbf{M}^{\mathrm{tb}}_{\mathrm{IB}}.
	\end{equation}
\end{subequations}

Now that both state-space models are created, the state-space model of the full system can be defined by using the ABH formulation as a feedback transfer function.
\section{Numerical demonstration of the proposed VABH}\label{sec:numerical features}
\subsection{Mechanical model}\label{sec:mechanical model section 3}
Euler-Bernoulli assumptions were used for beam modeling. One node comprises three degrees of freedom, namely the horizontal $u$ and vertical $v$ displacements and the rotation $\phi$. $400$ elements were employed for the uniform beam and $800$ for the tapered wedge beam. A uniform damping layer on both sides of either the cantilever beam or the tapered wedge beam was added to the model to dissipate the vibrational energy. The material and geometrical properties for each system are listed in Table~\ref{tab:parameters simu}. The damping matrix of each structure is built such that the considered resonance frequencies have all the same modal damping ratio $\xi$. Note that the residual thickness at the tip of the tapered wedge beam is very small, i.e., $\SI{0.6}{\micro\meter}$. Such a system would be very challenging to manufacture and would be very brittle. As mentioned in the Introduction, one of the main features of an ABH is its cut-on frequency, $f_{\mathrm{cut}}$. For the ABH in Table~\ref{tab:parameters simu}, we have~\cite{aklouche_scattering_2016}
\begin{equation}
	f_{\mathrm{cut}}=\frac{h_{0}}{2\pi L_{\mathrm{LABH}}^2}\sqrt{\frac{40E_{\mathrm{ABH}}}{12\rho_{\mathrm{ABH}}}}=\SI{9}{\hertz}.
\end{equation}
%\begin{figure}[H]
%	\centering
%	\includegraphics[scale=1]{body/images/Sketch_ABH_FEM_miror/beam_ABH.pdf}%,trim={5cm} {0} {0} {0},clip
%	\caption{Finite element model of the beam.}
%	\label{fig:FEM beam}
%\end{figure}

\begin{table}[H]
	\centering
	\begin{tabular}{cccc}
		\toprule
		\textbf{Parameter}&\textbf{Beam}&\textbf{ABH}&\textbf{Damping layer} \\
		\midrule
		Length&$L=\SI{1}{\meter}$&$L_{\mathrm{ABH}}=\SI{1}{\meter}$&$L_{\mathrm{d}}=\SI{1}{\meter}$ \\
		Width&$b=\SI{20}{\milli\meter}$&$b=\SI{20}{\milli\meter}$&$b=\SI{20}{\milli\meter}$ \\
		Thickness&$h_{0}=\SI{6}{\milli\meter}$&$m=2, \,x_{0}=\SI{10}{\milli\meter}$&$h_{\mathrm{d}}=\SI{0.8}{\milli\meter}$ \\
		Young modulus&$E_{\mathrm{b}}=\SI{210}{\giga\pascal}$&$E_{\mathrm{ABH}}=\SI{210}{\giga\pascal}$&$E_{\mathrm{d}}=\SI{5}{\giga\pascal}$ \\
		Density&$\rho_{\mathrm{b}}=\SI{7800}{\kilo\gram\per\meter\cubed}$&$\rho_{\mathrm{ABH}}=\SI{7800}{\kilo\gram\per\meter\cubed}$&$\rho_{\mathrm{d}}=\SI{920}{\kilo\gram\per\meter\cubed}$ \\
		Modal damping&$\xi_{\mathrm{b}}=0.05\%$ &$\xi_{\mathrm{ABH}}=0.05\%$&$\xi_{\mathrm{d}}=3\%$ \\
		\bottomrule
	\end{tabular}
	\caption{Parameters of the coupled system.}
	\label{tab:parameters simu}
\end{table}

\subsection{VABH definition}

\begin{figure}[H]
\centering
\includegraphics[scale=1]{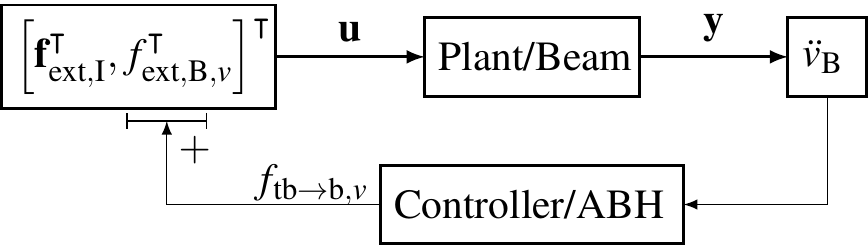}
\caption{Representation of the VABH as a feedback diagram.}
\label{fig:update SS VABH}
\end{figure}

The goal of this paper is to create a VABH according to Figure~\ref{fig:state space formulation}. However, it is difficult to measure experimentally the two accelerations and rotation at the interface and to apply the appropriate forces and torque. As a consequence, we consider here a truncated ABH effect, meaning that only the vertical acceleration $\ddot{v}_B$ is measured and only the vertical force $f_{tb\rightarrow b,v}$ is applied. Such a system is referred to as VABH in the remainder of this paper. Its feedback formulation is schematized in Figure~\ref{fig:update SS VABH}.

\subsection{Numerical results}

Three different simulations are considered for the purpose of comparison, namely one with no ABH, one with the full ABH (either mechanical or as a feedback transfer function), and one with the VABH. The features expected from a full mechanical ABH are described in different papers, see, e.g.,~\cite{pelat_acoustic_2020,tang_characterization_2016}. The system is excited at the tip of the beam. A damping layer is attached to the ABH system. For the case of the uniform beam with no ABH effect, the damping layer is attached to the beam itself.

\begin{figure}[H]
\centering
\begin{subfigure}{\textwidth}
\centering
    \includegraphics[scale=1]{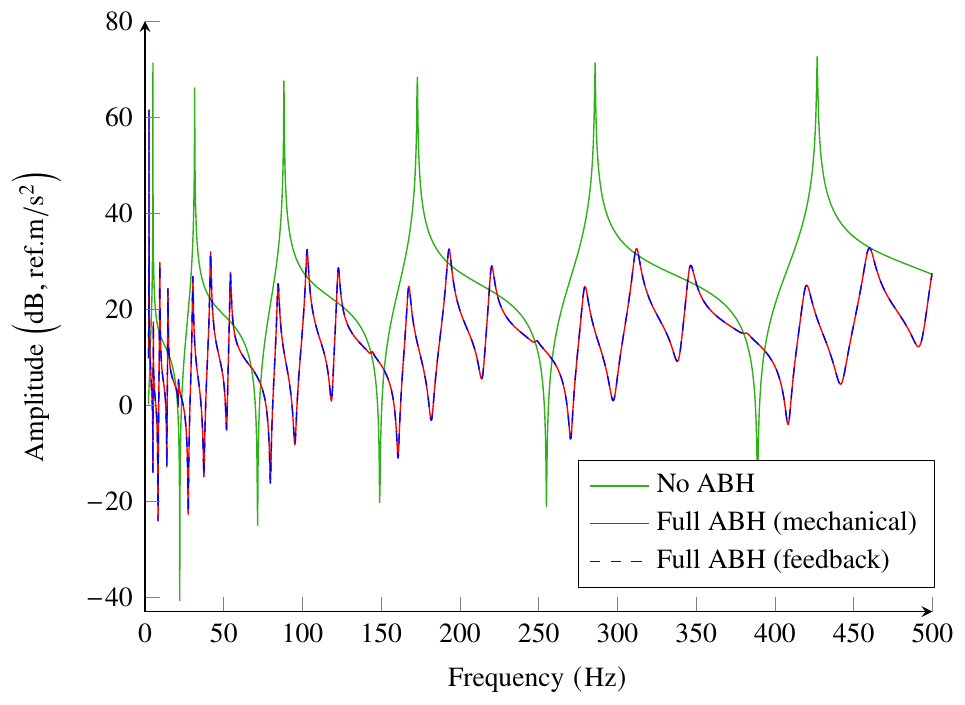}
    \caption{}
    \label{subfig:full ABH Frf}
\end{subfigure}%
			\par\bigskip % force a bit of vertical whitespace
%We now put the shape of the sector
\vspace{-0.4cm}
    \begin{subfigure}{\textwidth}
    \centering
    \includegraphics[scale=1]{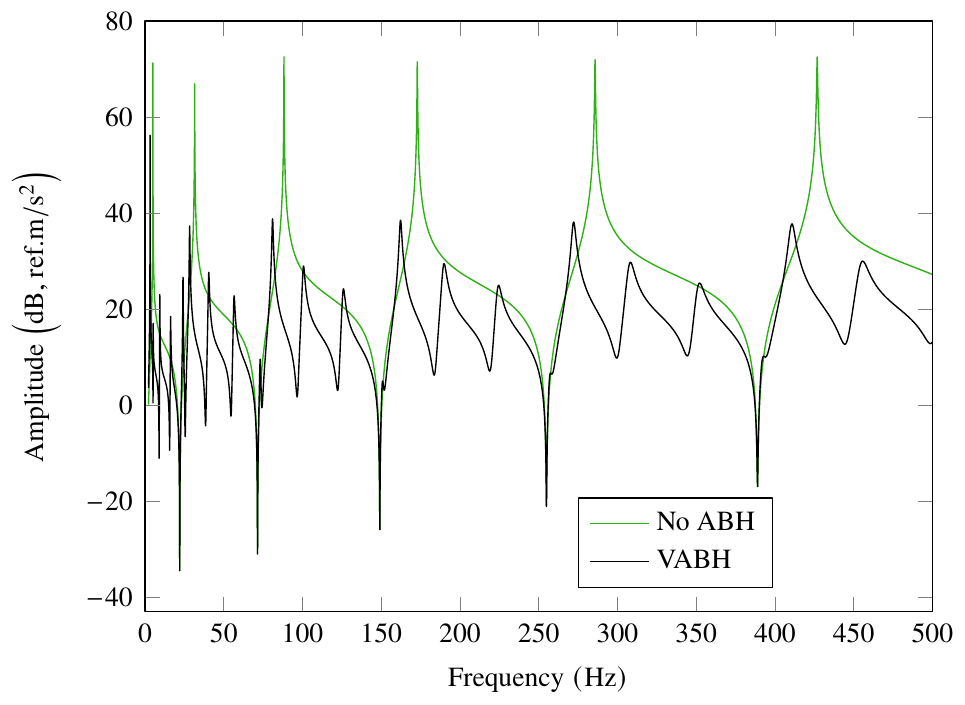}
    \caption{}
    \label{subfig:VABH Frf}
\end{subfigure}
    \caption{Frequency response functions measured at beam tip. (a) No ABH and full ABH; (b) no ABH and VABH. The cut-on frequency is represented with the dashed line.}
	\label{fig:frf caract}
\end{figure}

Figure~\ref{fig:frf caract} presents the frequency response functions (FRFs) measured at beam tip. Figure~\ref{fig:frf caract}(a) confirms the complete equivalence between the mechanical ABH and the ABH defined through the feedback formulation in Figure~\ref{fig:state space formulation}. As expected, the vibration of the beam equipped with the mechanical ABH is significantly reduced for frequencies beyond $f_{\mathrm{cut}}$
compared to the case without ABH. 
Interestingly, the FRF of the VABH beam in Figure~\ref{fig:frf caract}(b) exhibits performance comparable to that of the mechanical ABH. We stress that this result was obtained for different measurement locations and for different mechanical systems, but this is not further discussed herein for conciseness. 

\begin{figure}[H]
	\centering
	\includegraphics[scale=1]{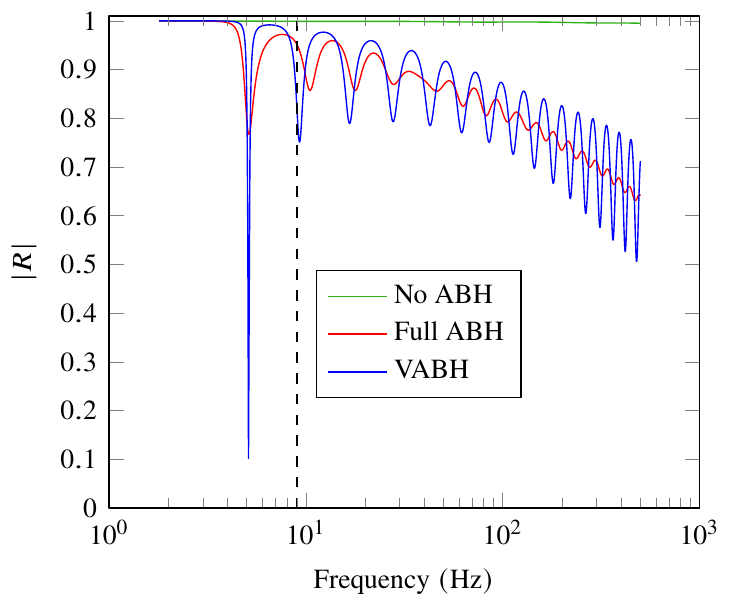}
	\caption{Reflection coefficient $\left|R\right|$. The cut-on frequency is represented with the dashed line.}
	\label{fig:refletion caract}
\end{figure}

To better highlight the possible differences between the VABH and the standard ABH, several indicators were calculated. Figure~\ref{fig:refletion caract} displays the reflection coefficient $\left|R\right|$ defined based on the experimental technique introduced in~\cite{denis_measurement_2015}. It is equal to one for the uniform beam, because the propagating wave is totally reflected at the boundary. For the full ABH, the reflective coefficient fluctuates along the frequency range as discussed in~\cite{denis_measurement_2015}. Moreover, as the frequency increases, its value tends to decrease as predicted by the geometrical acoustics method~\cite{krylov_acoustic_2004}. The same trends are observed for the VABH. However, the fluctuation is more pronounced in the case of the VABH. 

\begin{figure}[H]
	\centering
	\includegraphics[scale=1]{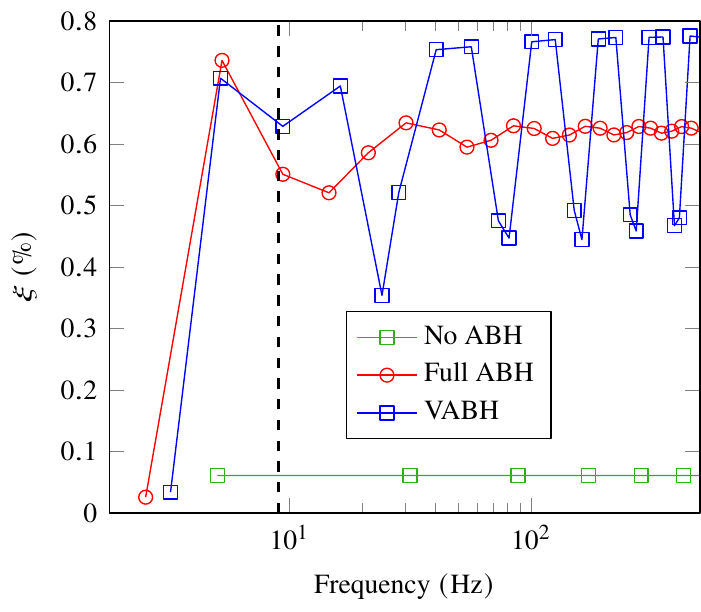}
	\caption{Modal damping ratio $\xi$. The modes are represented with markers. The cut-on frequency is represented with the dashed line.}
	\label{fig:damping caract}
\end{figure}

Figure~\ref{fig:damping caract} exhibits the modal damping ratio $\xi$ for the different modes of the considered systems. For the uniform beam upon which a damping layer is attached, the overall damping is around $0.06\%$. The first mode of both the full ABH and VABH has a frequency around $\SI{3}{\hertz}$, i.e., below the cut-on frequency. This explains why it exhibits a lower damping ratio around $0.03\%$. For the higher modes, the damping ratio increases substantially and fluctuates around a value of $0.6\%$. This increase is also visible for the second mode which is located below the cut-on frequency, but this is consistent with previous observations in the technical literature~\cite{pelat_acoustic_2020}.

%\begin{figure}[H]
%	\centering
%	\frame{\includegraphics[scale=1]{body/images/caracteristiques_Frf_poutre_bis-src/caracteristiques_Frf_poutre_bis.pdf}}
%	\caption{Frequency response functions of the three mechanical systems. The cut-on frequency is localized with the black dashed line.}
%	\label{fig:frf caract}
%\end{figure}

\begin{figure}[H]
	\centering
	\includegraphics[scale=1]{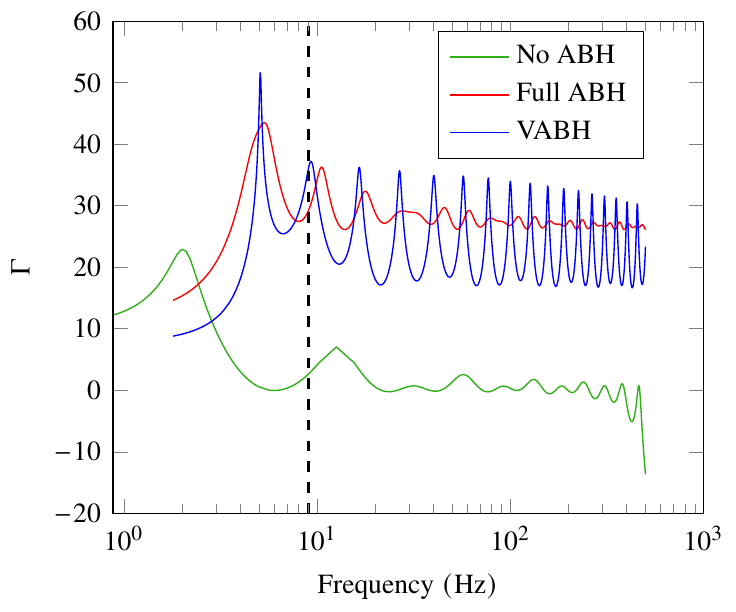}
	\caption{Energy ratio $\Gamma$. The cut-on frequency is represented with the dashed line.}
	\label{fig:energy caract}
\end{figure}

Figure~\ref{fig:energy caract} compares the energy ratio defined as
\begin{equation}
	\Gamma=10\log\left(\sqrt{\frac{N_{\mathrm{b}}\mathlarger{\sum}_{i=N_{\mathrm{b}}+1}^{N_{\mathrm{b}}+N_{\mathrm{tb}}}v_{i}^{2}}{N_{\mathrm{tb}}\mathlarger{\sum}_{i=1}^{N_{\mathrm{b}}}v_{i}^{2}}}\right).
\end{equation}
It is equal to the ratio of the mean quadratic velocity of the tapered wedge part with respect to that of the uniform beam; it indicates where the energy density is located. For the uniform beam with no ABH, a longer uniform beam is considered with a length equal to $L+L_{\mathrm{ABH}}$. For this system, the energy ratio is defined as the quadratic mean velocity of the nodes contained between $\left[L, L+L_{\mathrm{ABH}}\right]$ with the nodes contained in $\left[0,L\right]$. For structures equipped with an ABH, a positive energy ratio is systematically observed. It means that, over the entire frequency range, the energy in the ABH is greater than in the uniform beam. Between $\SI{5}{\hertz}$ and $\SI{1000}{\hertz}$, the energy ratio of mechanical systems with an ABH is always greater than that of the uniform beam. These results are similar to those in~\cite{tang_characterization_2016}. As for Figures~\ref{fig:refletion caract} and ~\ref{fig:damping caract}, the VABH presents the same features as the standard ABH but with higher amplitude fluctuations. 

In summary, despite the fact that it only considers the vertical acceleration and the vertical interface force, the VABH exhibits performance similar to that of the full ABH system.

%For the current case, assembling the uniform beam with a damping layer creates a system with additional damping (around $\SI{6e-4}{}$). When assembling the uniform and tapered wedge beams, its damping coefficient for the first mode becomes lower than the one of either subsystem. Attaching the damping layer to the tapered wedge beam allows to increase its value up to $\SI{3e-4}{}$.

\section{Experimental considerations}\label{sec:final app}

\subsection{Experimental setup}

\begin{figure}[H]
	\centering
	\includegraphics[height=4cm]{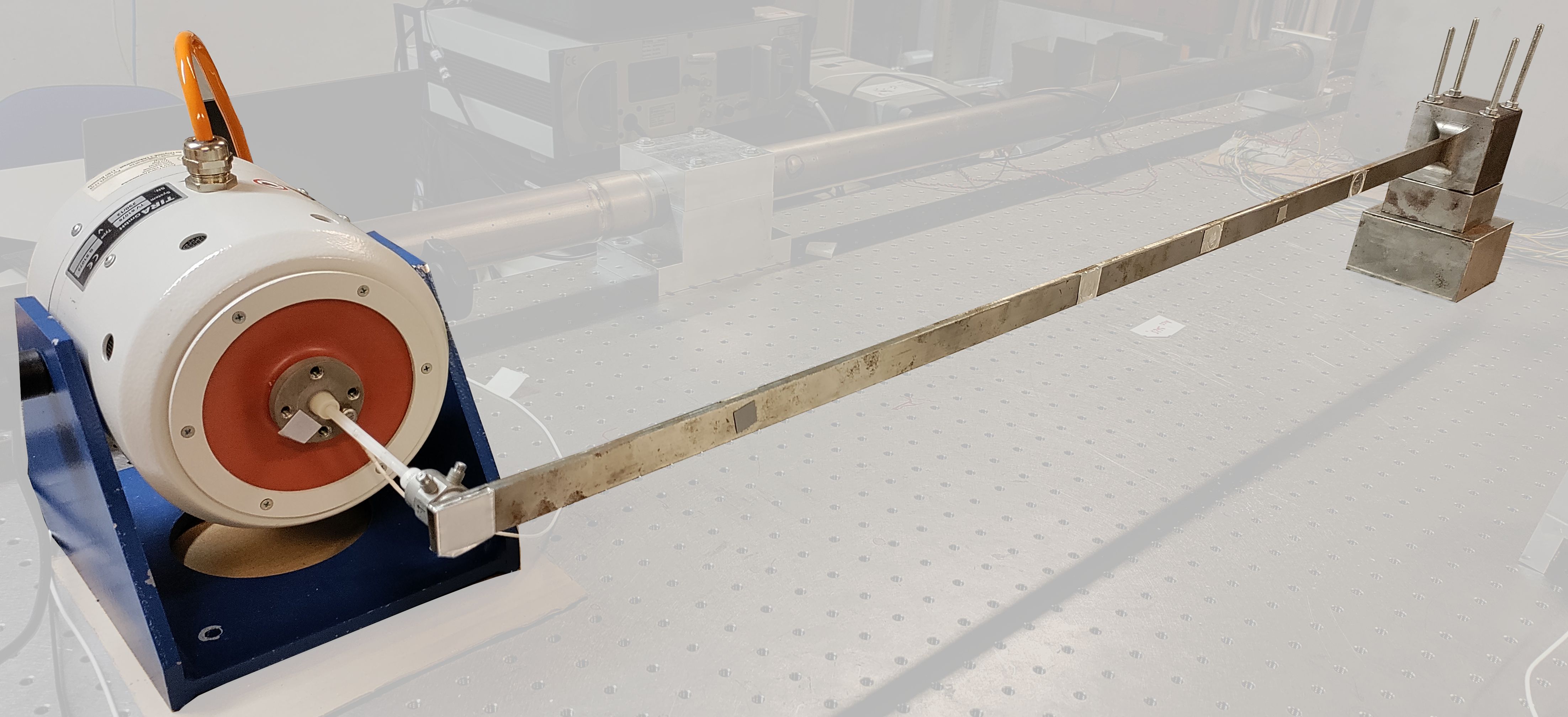}
	\caption{Experimental setup}
	\label{fig:experimental picture}
\end{figure}

The experimental setup, previously employed in~\cite{abeloos_stepped_2021}, is depicted in Figure~\ref{fig:experimental picture}. The uniform rectangular steel cantilever beam has the same parameters as those in Table~\ref{tab:parameters simu}. To avoid friction and contact nonlinearities, the base and the beam were manufactured from one steel block, and the base was bolted to the ground.  The system was excited and controlled using the same electrodynamic shaker (TIRA TV51075). A metallic cap with an impedance head of $\SI{60}{\gram}$ was attached to the beam. 

The excitation signal and the VABH control law were generated with a real-time controller (RTC) DSpace MicroLabBox with a sampling frequency of $\SI{100}{\kilo\hertz}$. The excitation was a multi-sine function in the $\left[\SI{0.1}{\hertz},\SI{500}{\hertz}\right]$ frequency interval with an amplitude of $\SI{0.01}{\volt}$. The gain between the command in volt and the applied force is $\SI{160}{\newton\per\volt}$. The controller, initially written as a state-space system, was transformed into a single input single output (SISO) state-space model. The system was subsequently transformed into a zero-pole-gain function and finally discretized with Tustin's method for implementation into the RTC. Ideally, the controller must be proper and stable to be implemented. To ensure stability, a collocated plant transfer function~\cite{preumont_collocated_2011} is almost mandatory, i.e., the sensor and actuator of the feedback function must be located at the same position. The electrodynamic shaker was placed as close as possible from the beam edge  (i.e., at $\SI{20}{\milli\meter}$ of the free tip) with an accelerometer glued to it. Overall, the effective moving mass (magnetic coil+accelerometer) is equal to $\SI{150}{\gram}$. To reject undesirable high-frequency shaker dynamics and to have a proper transfer function, a low pass filter 
\begin{equation}\label{eq:low pass filter}
	h_{\mathrm{LP}}=\dfrac{\omega_{\mathrm{LP}}^2}{s^2+\omega_{\mathrm{LP}}s+\omega_{\mathrm{LP}}^2},
\end{equation}
was applied to the controller where $\omega_{\mathrm{LP}}$ is the low-pass cut-on frequency and $s$ is Laplace's variable. The low-pass frequency was set to $\SI{800}{\hertz}$.

\subsection{Transfer function of the plant}
%\subsubsection{Case of the plant: update of the beam finite element model}

\begin{figure}[H]
	\centering
	\includegraphics[scale=1]{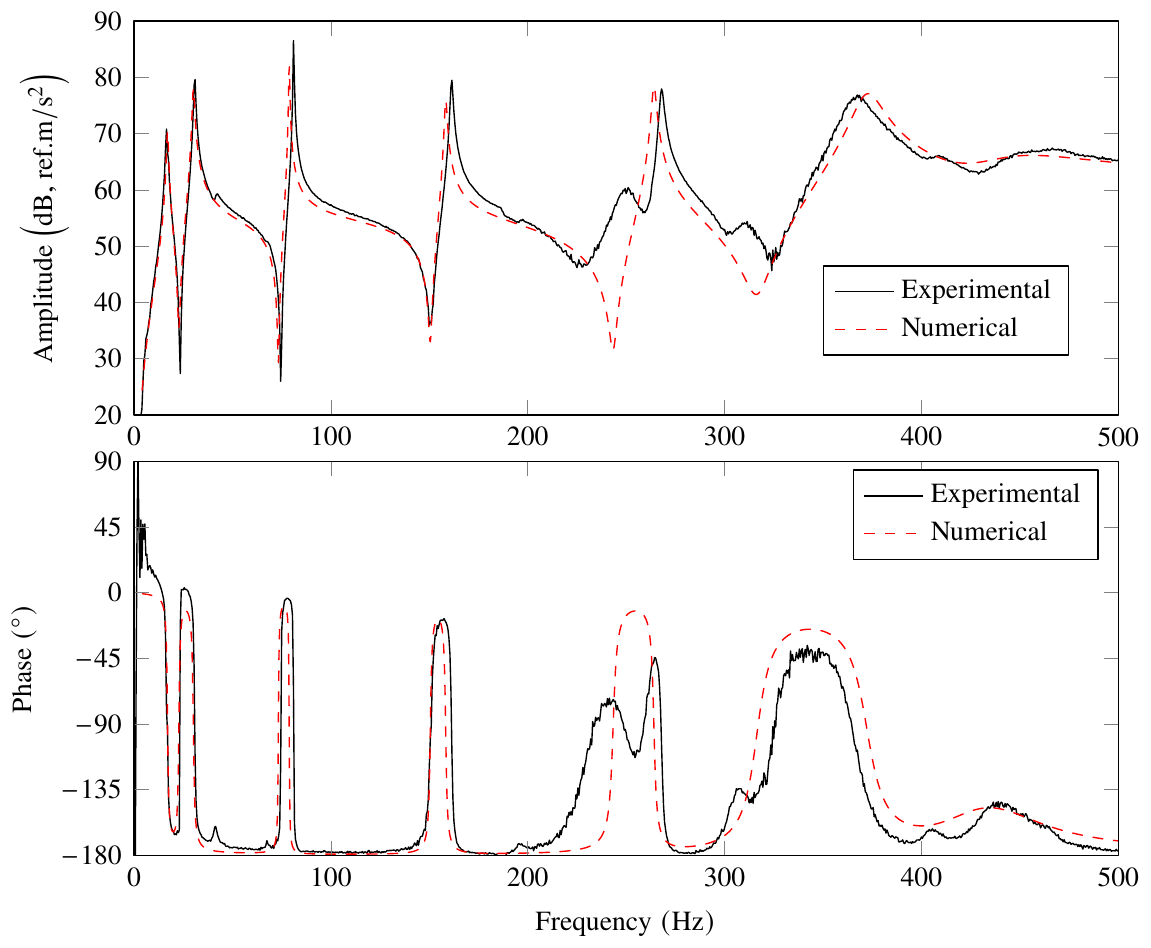}
	\caption{Bode plots of the uniform beam with no VABH.}
	\label{fig:comparison experimental numerique}
\end{figure}

The Bode plot of the experimental setup is provided in Figure~\ref{fig:comparison experimental numerique}. The transfer function presents collocated features, i.e., a pole-zero alternation and a phase bounded between $\SI{-180}{\degree}$ and $\SI{0}{\degree}$ (except at very low frequencies where the accelerometer is no longer measuring accurately). We note that both modes 3 and 10 are in the vicinity of high amplitude modes, which is most likely due to a coupling between the uniform beam and the stinger's dynamics. Modes 6 and 8 present a low amplitude vibration because they correspond to torsional modes. The identified resonance frequencies $f_{\mathrm{res}}$ and modal damping coefficients $\xi$ are given in Table~\ref{tab:experimental damping and frequencies}. The values are different than those in~\cite{abeloos_stepped_2021} because the measured FRF includes the dynamics of both the stinger and the shaker.

\begin{figure}[H]
	\centering
	\includegraphics[scale=1]{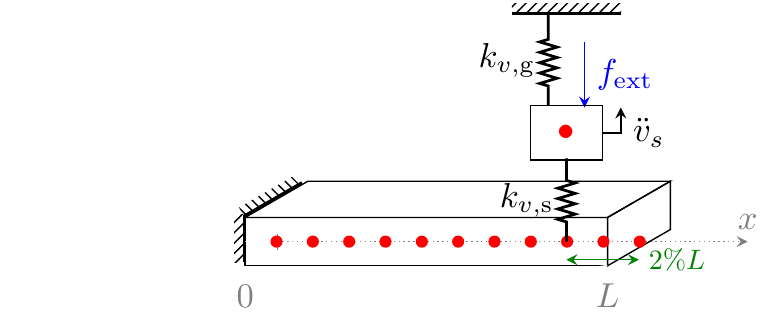}%,trim={5cm} {0} {0} {0},clip
	\caption{Upgraded finite element model of the beam.}
	\label{fig:FEM beam expe}
\end{figure}

The finite element model was upgraded to account for the presence of the exciting and measurements devices, as illustrated in Figure~\ref{fig:FEM beam expe}. The transversal mass located at $\SI{20}{\milli\meter}$ of the beam tip was updated to account for the impedance head. Besides, a transversal degree of freedom was added to account for the accelerometer and the shaker. It is linked to ground with a stiffness $k_{v,\mathrm{g}}$ representing the shaker membrane's stiffness and to  the beam by $k_{v,\mathrm{s}}$ representing the stinger's stiffness. Measuring the transfer function of the shaker detached from the stinger enabled us to determine that $k_{v,\mathrm{g}}=\SI{6.5e3}{\newton\per\meter}$. The value of the stiffness $k_{v,\mathrm{s}}$, $\SI{4e5}{\newton\per\meter}$, was obtained by minimizing the differences between the resonance frequencies of the experimental setup and of the numerical model. In the optimization process, modes 3, 6, 8 and 11 were not retained because they could not be obtained with the finite element model. Overall, an excellent agreement between the experimental and numerical transfer functions can be observed in Figure~\ref{fig:comparison experimental numerique}.

\begin{table}[H]
	\centering
	\begin{tabular}{cccccccccccc}
		\toprule
		\textbf{Modes}&1&2&3&4&5&6&7&8&9&10&11 \\
		\midrule
		\textbf{$f_{\mathrm{res}}$} in $\left(\SI{}{\hertz}\right)$ &$16.5$&$31$&$42.5$&$81$&$161.5$&$249.5$&$268$&$311.5$& $367$&$408.5$&$458$ \\
		\textbf{$\xi \left(\%\right)$} &$\SI{2.65}{}$&$\SI{1.79}{}$&NA&$\SI{0.28}{}$&$\SI{0.41}{}$&$\SI{2.2}{}$&$\SI{0.41}{}$&NA &$\SI{1.91}{}$& NA &$\SI{7.41}{}$\\
		\bottomrule
	\end{tabular}
	\caption{Resonance frequencies and modal damping coefficients of the experimental setup. NA means that the value could not be identified.}
	\label{tab:experimental damping and frequencies}
\end{table}

%\begin{figure}[H]
%	\centering
%	\includegraphics[scale=1]{body/images/comparison_expe_num_plant-src/comparison_expe_num_plant.pdf}
%	\caption{Comparison of the frequency response functions between the experiment and numerical simulations.}
%	\label{fig:comparison experimental numerique}
%\end{figure}

\subsection{Open loop system and stability margins}\label{subsec:stability margins}
%Besides the measures taken in Section~\ref{subsec:proper and stable controller}, stability margins of the open-loop transfer function~\cite{preumont_stability_2011} need to be verified before running an experiment.

The response of the system with the feedback is called closed-loop transfer function. For a SISO system, it reads
\begin{equation}\label{eq:close loop TF}
    T(s)=\dfrac{P(s)}{1+P(s)Q(s)},
\end{equation}
where $P$ and $Q$ are the transfer functions of the plant and of the controller, respectively. To ensure stability, the roots of the denominator must all have a negative real part. If there exists a frequency $\omega$ such that $1+P(j\omega)Q(j\omega)=0$ (with $j$ the unit imaginary number), the closed-loop system possesses a pair of complex conjugate poles on the imaginary axis, i.e., the system is marginally stable. Any perturbation to the nominal system may change their position to either side of the imaginary axis and is thus susceptible to make the controlled system unstable. This happens when the product (also called open-loop transfer function) $P(j\omega)Q(j\omega)=-1$, or equivalently
\begin{equation}
    \left|PQ\right|= 1,\hspace{1em}\text{and}\hspace{1em}\mathrm{arg}\left(PQ\right)=\pm\SI{180}{\degree},
    \label{eq:openLoopCriticalPoint}
\end{equation}
where $\mathrm{arg}$ is the argument of a complex number. 

Stability margins are commonly used to quantify how far the open-loop transfer function is from the conditions of Equation~\ref{eq:openLoopCriticalPoint}.
The gain margins correspond to the value of $0-\left|PQ\right|$ when $\mathrm{arg}\left(PQ\right)=\pm\SI{180}{\degree}$ and indicate the factor by which the gain in the loop can be increased to make the system marginally stable. Similarly, the phase margins correspond to the value of $\pm180-\mathrm{arg}\left(PQ\right)$ when $\left|PQ\right|\neq 1$ and indicate how robust the system is to delays in the loop. In practice, $\SI{40}{}-\SI{50}{\degree}$ of stability margins is acceptable~\cite{franklin2002feedback}. Gain margins should be positive and greater than a few $\SI{}{\decibel}$.

% This is achieved by multiplying the transfer function of the plant with that of the controller.

\begin{figure}[H]
	\centering
	\includegraphics[scale=1]{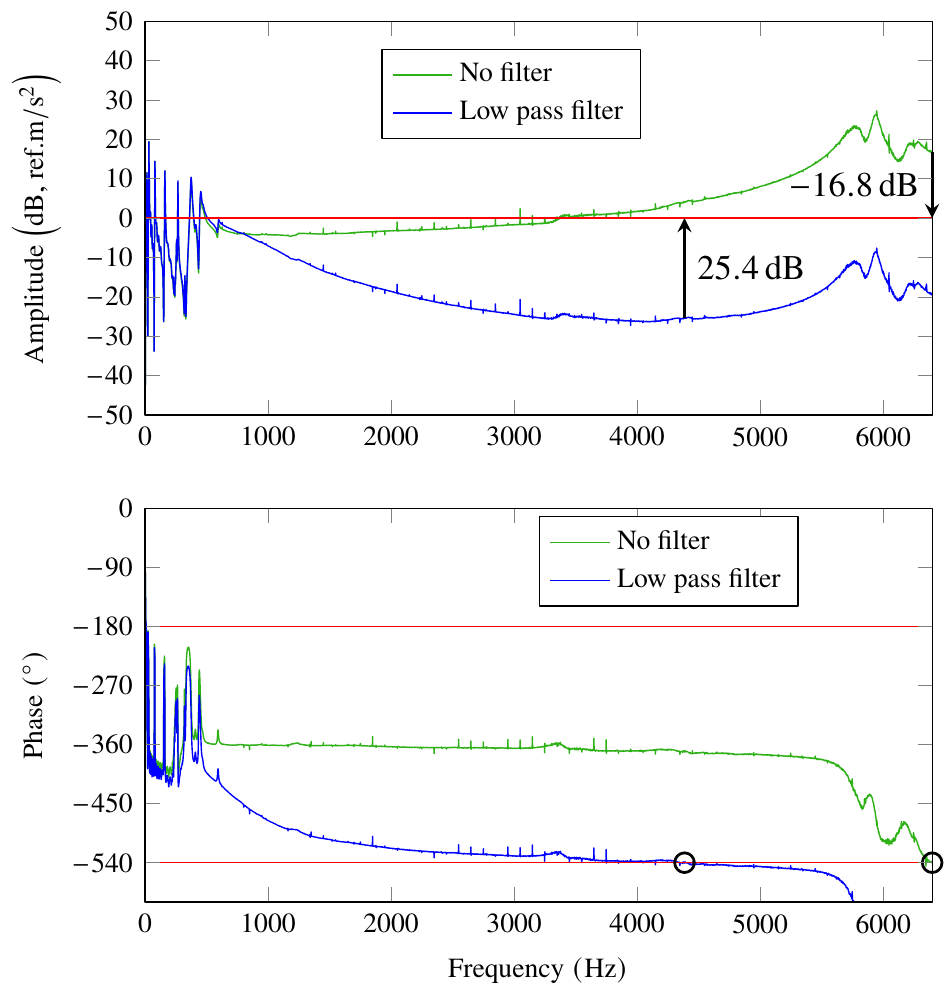}
	\caption{Stability margins of the open-loop system. Only one gain margin is given as an example.}
	\label{fig:stability margins full}
\end{figure}

\begin{figure}[H]
	\centering
	\includegraphics[scale=1]{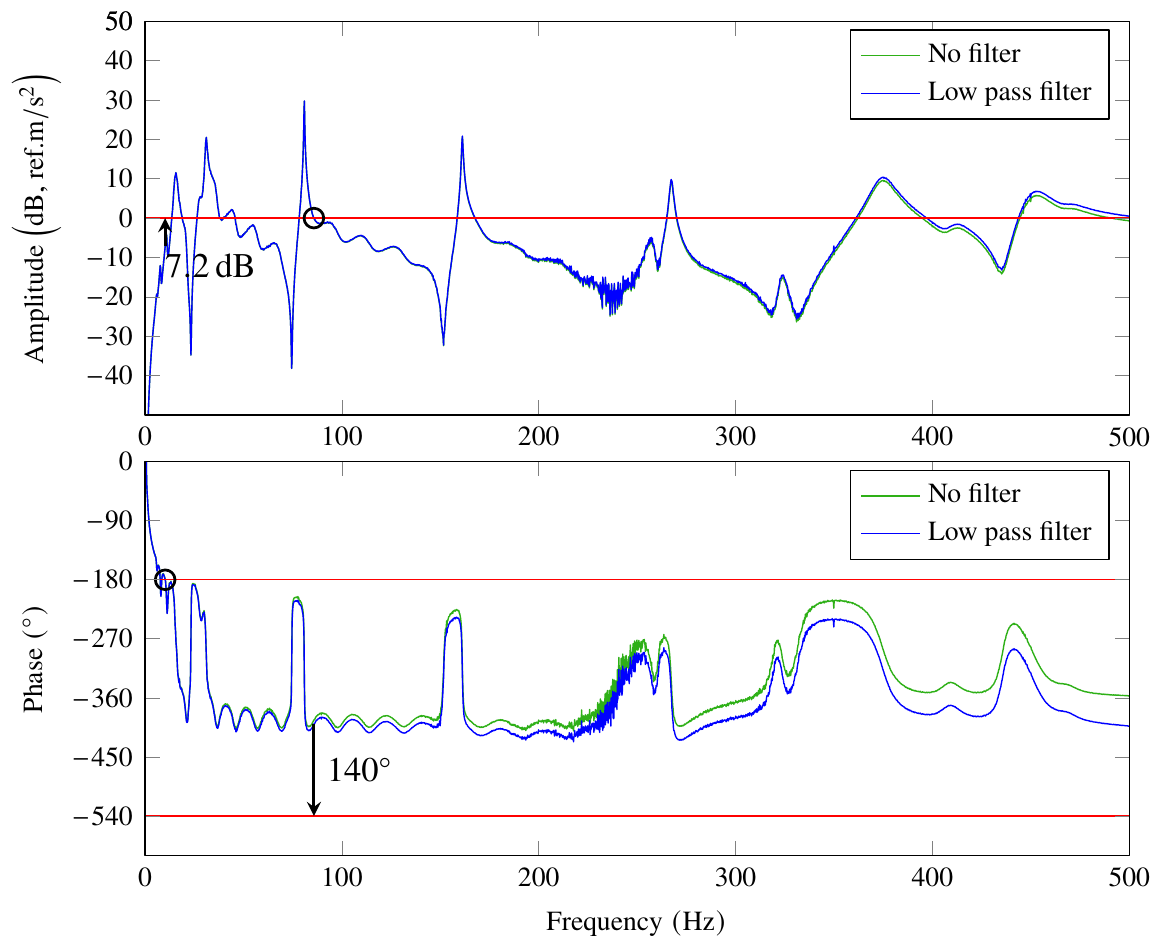}
	\caption{Close-up of Figure~\ref{fig:stability margins full} in the frequency range of interest.}
	\label{fig:stability margins zoom}
\end{figure}

Figure~\ref{fig:stability margins full} presents the stability margins for the controller given in Section~\ref{sec:numerical features}. We note that, instead of inserting a damping layer in the VABH, its  modal damping $\xi_{\mathrm{ABH}}$ was rather increased up to $5\%$. To underline the importance of the low pass filter, two open-loop transfer functions are compared that is one with the filter and one without. Beyond the shaker bandwidth around $\SI{6}{\kilo\hertz}$, the modes of the shaker are visible. Without the low pass filter, negative gain margin is obtained and thus the closed-loop system would be unstable. However, with the filter, the effect of the shaker is lessened, and stability is ensured. A close-up is shown in Figure~\ref{fig:stability margins zoom}. We note that phase margins close to $\SI{-180}{\degree}$ do not present a risk to the stability of the system. Indeed, time delays induced by the processor tend to lower the phase of the system and thus increase these phase margins. Similarly, the gain margin of $\SI{7.2}{\decibel}$ at $\SI{10}{\hertz}$ is expected to be higher during the experiment.

\subsection{Closed-loop system}\label{sec:close loop}

The fundamental difference between the VABH in Section~\ref{sec:mechanical model section 3} and the experimental VABH comes from the size of the ABH's dynamical model which needs to be reduced substantially. Indeed, the very fine discretization used for the tapered wedge beam model would result in a controller of very high order (number of poles and zeros), which could, in turn, not be handled by the RTC's processor. 
To preserve the ABH's dynamics and make it digital through the RTC, a Craig-Bampton reduction method~\cite{craig_coupling_1968} was used on the tapered wedge beam. This substructuring approach partitions the degrees of freedom into master and slave degrees of freedom. The former are kept as physical nodes whereas the latter are reduced into $N_{\mathrm{CB}}$ generalized control coordinates. In our case, the boundary nodes were kept as master nodes to perform the assembly and $N_{\mathrm{CB}}=20$ was the maximum value allowed by the RTC.

\begin{figure}[H]
\centering
\includegraphics[scale=1]{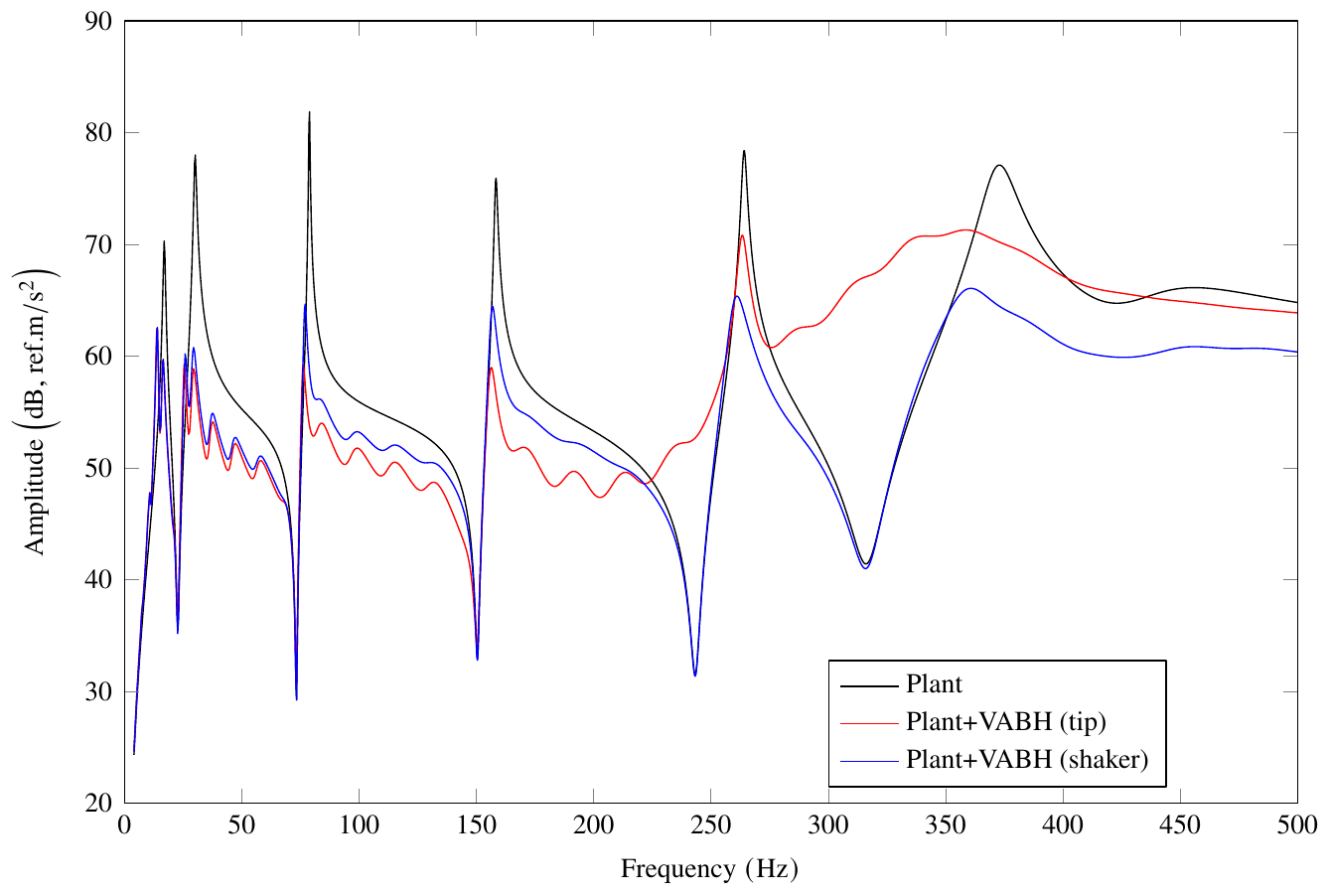}
\caption{Influence of ABH location on the FRF (numerical result).}
\label{fig:full influence}
\end{figure}

The impact of this model reduction together with the influence of the position of the VABH and of the low-pass filter was studied numerically where the controller corresponds to the one used in Section~\ref{subsec:stability margins}. Figure~\ref{fig:full influence} compares a VABH applied directly at the tip (in red) with one located at the shaker's position (in blue). For the first four modes, the VABH at the tip presents slightly better vibration reduction. For frequencies greater than $\SI{280}{\hertz}$, the VABH placed at the shaker exhibits better performance. Surprinsingly, the VABH at the tip shows greater vibration amplitudes than those of the plant in the range $\left[\SI{300}{\hertz},\SI{370}{\hertz}\right]$. This result comes from the stiffness of the stinger $k_{v,\mathrm{s}}$, which, combined with the ABH placed at the tip, creates a number of resonant frequencies in $\left[\SI{300}{\hertz},\SI{370}{\hertz}\right]$. We note that the modification of the position of an ABH was studied in detail in~\cite{zhou_resonant_2018}.

\begin{figure}[H]
\centering
\includegraphics[scale=1]{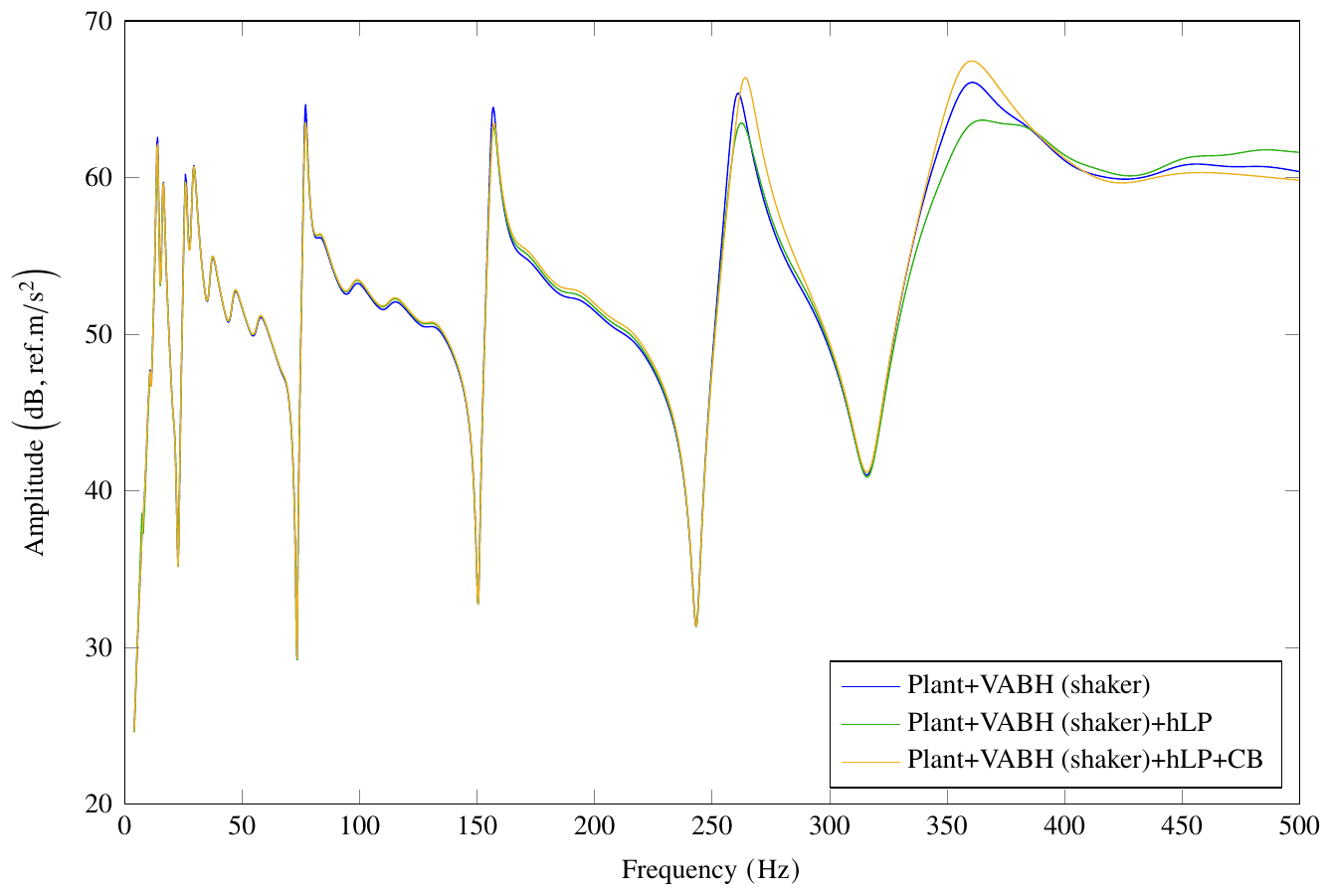}
\caption{Influence of the filtering and model reduction on the FRF (numerical result).}
\label{fig:full filtering}
\end{figure}

Figure~\ref{fig:full filtering} plots the influence of the low pass-filter on the FRF (blue vs. green). For low-frequency modes, $\omega_{\mathrm{LP}}$ is high enough so that the filter has no influence. For the modes at $\SI{268}{\hertz}$ and $\SI{367}{\hertz}$, vibration reduction is slightly enhanced with the filter. 

Finally, Figure~\ref{fig:full filtering} presents the influence of the reduction strategy (green vs. yellow). Except for small discrepancies for the higher modes, keeping $20$ modes in the basis does not modify the FRF to a great extent.

\section{Experimental performances of the VABH}\label{sec:performances ABH}
\subsection{Nominal VABH}\label{sec:nominal}
\begin{figure}[H]
	\centering
	\includegraphics[scale=1]{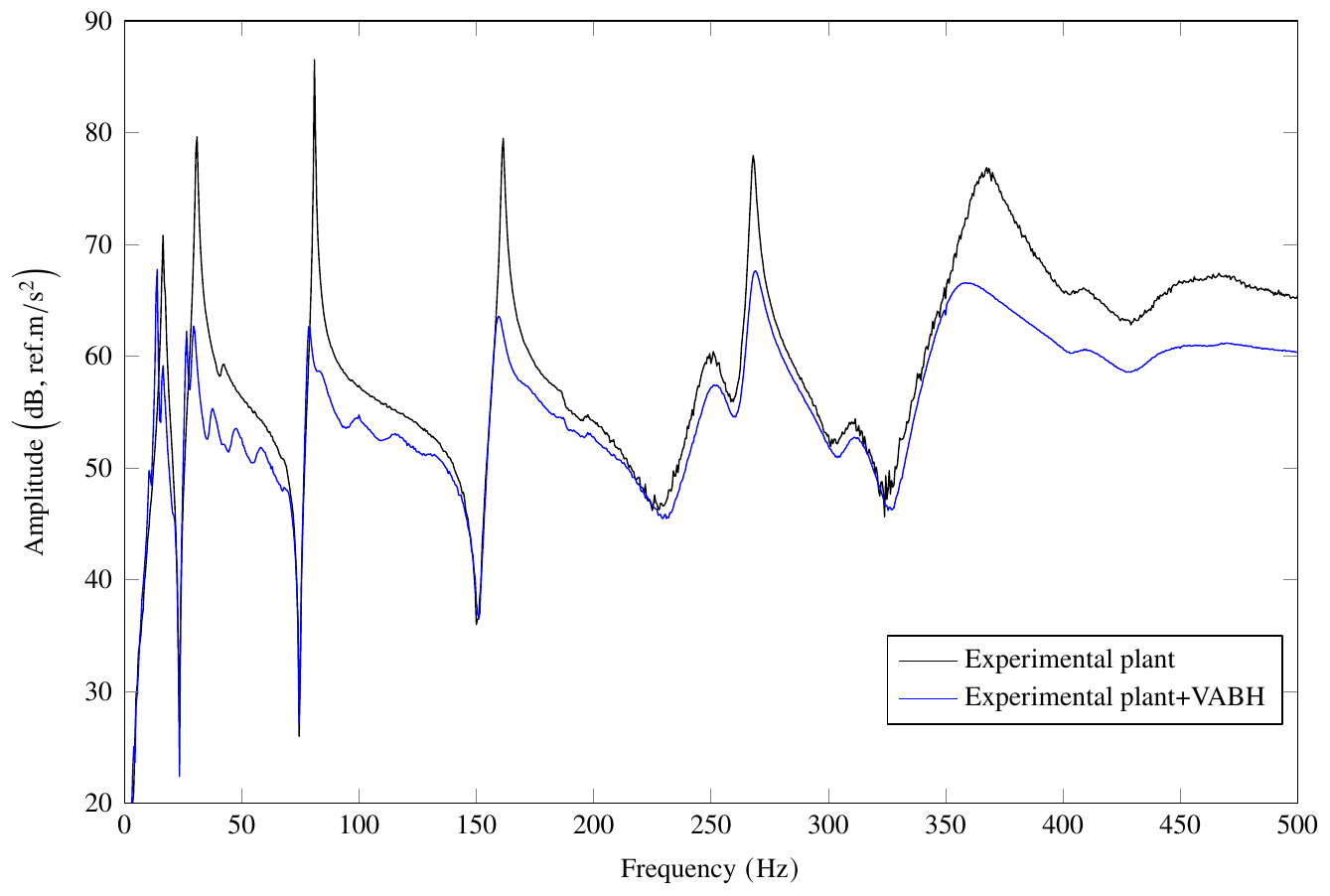}
	\caption{Experimental FRF of the plant and of the plant+VABH system.}
	\label{sec:comp expe plant}
\end{figure}

\begin{figure}[H]
	\centering
	\includegraphics[scale=1]{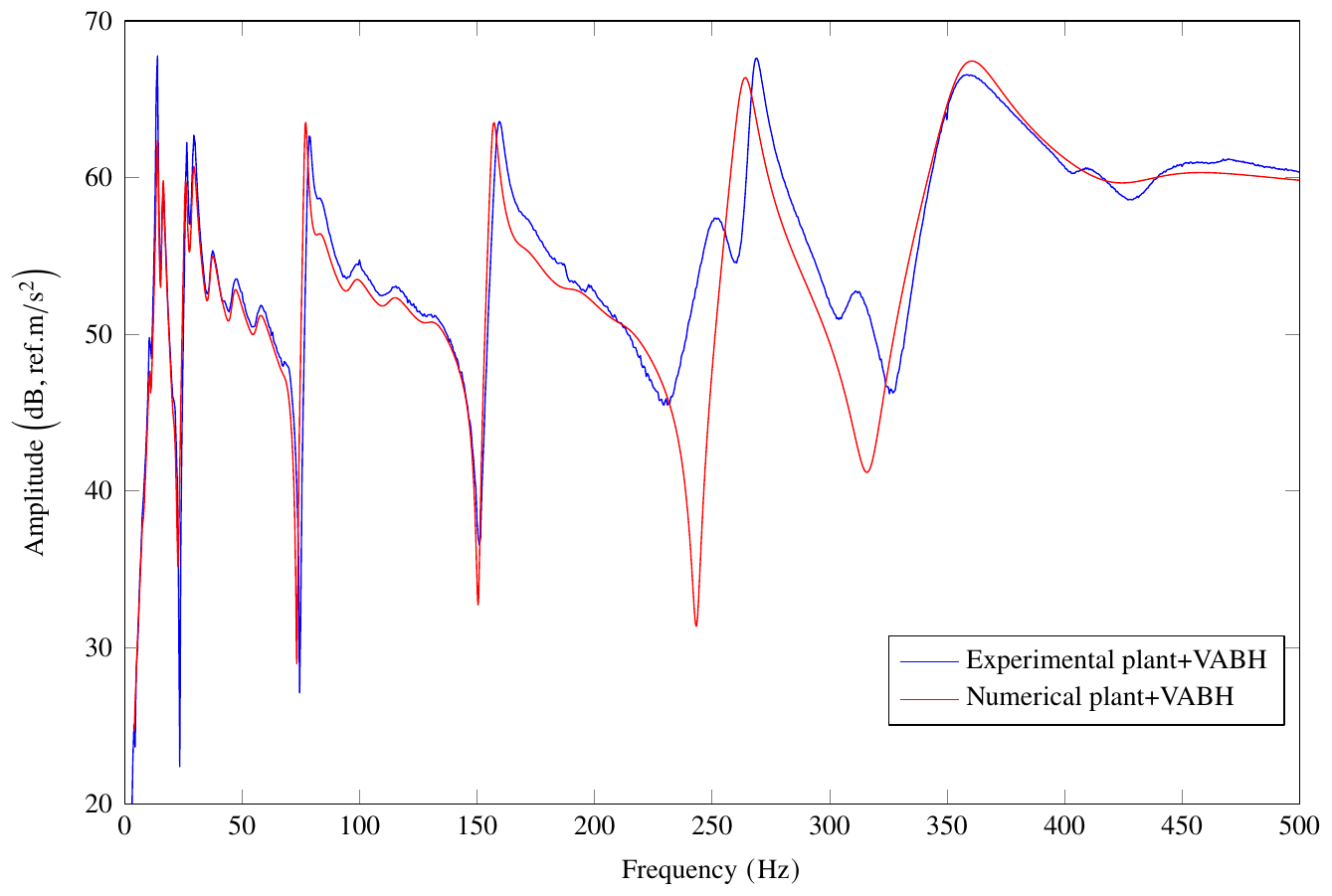}
	\caption{Comparison between experimental and numerical FRFs of the plant+VABH system.}
	\label{sec:comp expe num}
\end{figure}

The plant+VABH system is now considered experimentally. The nominal VABH is described by the parameters in Table~\ref{tab:parameters simu} with $\xi_{\mathrm{ABH}}=5\%$. Figure~\ref{sec:comp expe plant} presents, for the first time, the implementation of a VABH on an experimental setup. The attenuation provided by the VABH is given in Table~\ref{tab:attenuation} for different resonance frequencies of the plant. With the exception of the first mode, all modes are significantly damped with the greatest attenuation amounting to $\SI{23.9}{\decibel}$. Thus, excellent attenuation performance is offered by the VABH. These results are also well-predicted by the numerical model, as exemplified in Figure~\ref{sec:comp expe num}.

\begin{table}[H]
	\centering
	\begin{tabular}{cccccccc}
		\toprule
		\textbf{Experimental plant modes}&1&2&4&5&7&9&11 \\
		\midrule
		\textbf{Experimental attenuation}&3&17.4&23.9&15.9&10.3&10.3&6.2 \\
		\textbf{Numerical attenuation}&8&17.1&15.4&11.8&12.1&9.7&5.8 \\
		\bottomrule
	\end{tabular}
	\caption{Attenuation (in $\SI{}{\decibel}$) of the plant's resonance peaks using the VABH (20 modes). Modes 3, 6, 8, and 10 are excluded, because they correspond to low-amplitude vibration modes.}
	\label{tab:attenuation}
\end{table}

\subsection{Parametric studies}

Because the VABH relies entirely on a microcontroller, changing parameters such as the length or the material of the ABH can easily be carried out. As it will be shown, better performance compared to that of the nominal ABH can be obtained.

\begin{figure}[H]
\centering
	\includegraphics[scale=1]{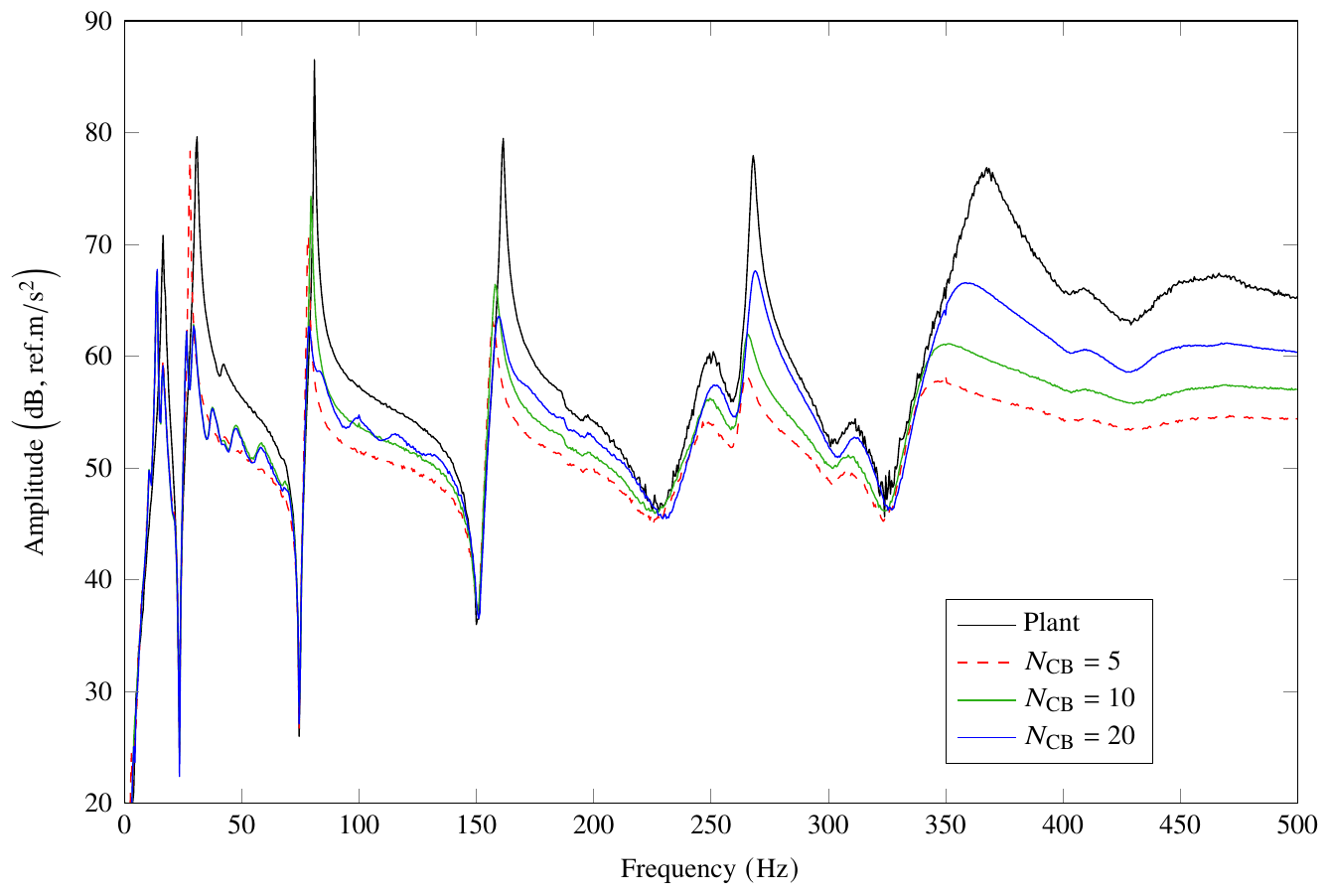}
	\caption{Influence of the number of Craig-Bampton modes kept during model reduction on ABH performance.}
	\label{subfig: CB expe}
\end{figure}

The influence of the number of modes kept during the Craig-Bampton reduction is first studied in Figure~\ref{subfig: CB expe}. When the reduction is performed with a smaller number of modes, the VABH performance decreases at low frequencies and increases for higher-frequency modes. This latter observation was not expected and should be interpreted with care because the dynamics of the ABH is not well described at high frequencies. In the next experiments, the controller is always implemented with 20 modes.

\begin{figure}[H]
	\centering
	\includegraphics[scale=1]{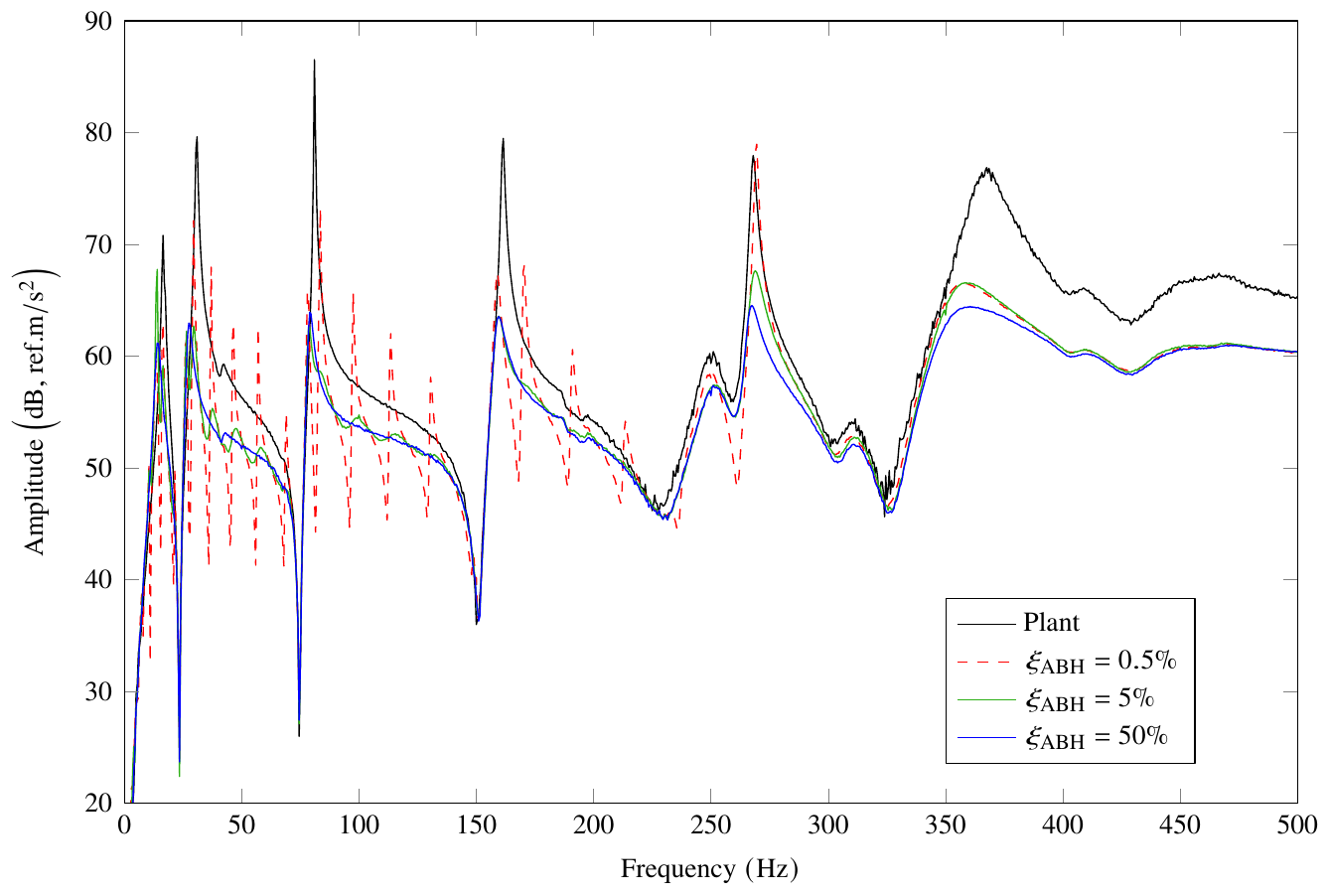}
	\caption{Influence of the modal damping on ABH performance.}
	\label{fig:influence damping}
\end{figure}

Three values of the damping ratio $\xi_{\mathrm{ABH}}$, namely $0.5\%$, $5\%$, and $50\%$ are considered in Figure~\ref{fig:influence damping}. When $\xi_{\mathrm{ABH}}=0.5\%$, the VABH is not very effective as all resonance peaks present fairly high amplitudes. Setting $\xi_{\mathrm{ABH}}$ to $=50\%$ reduces substantially the amplitudes of all resonance peaks. A $\SI{10}{\decibel}$ attenuation is now obtained for the first resonance peak of the plant.

\begin{figure}[H]
	\centering
	\includegraphics[scale=1]{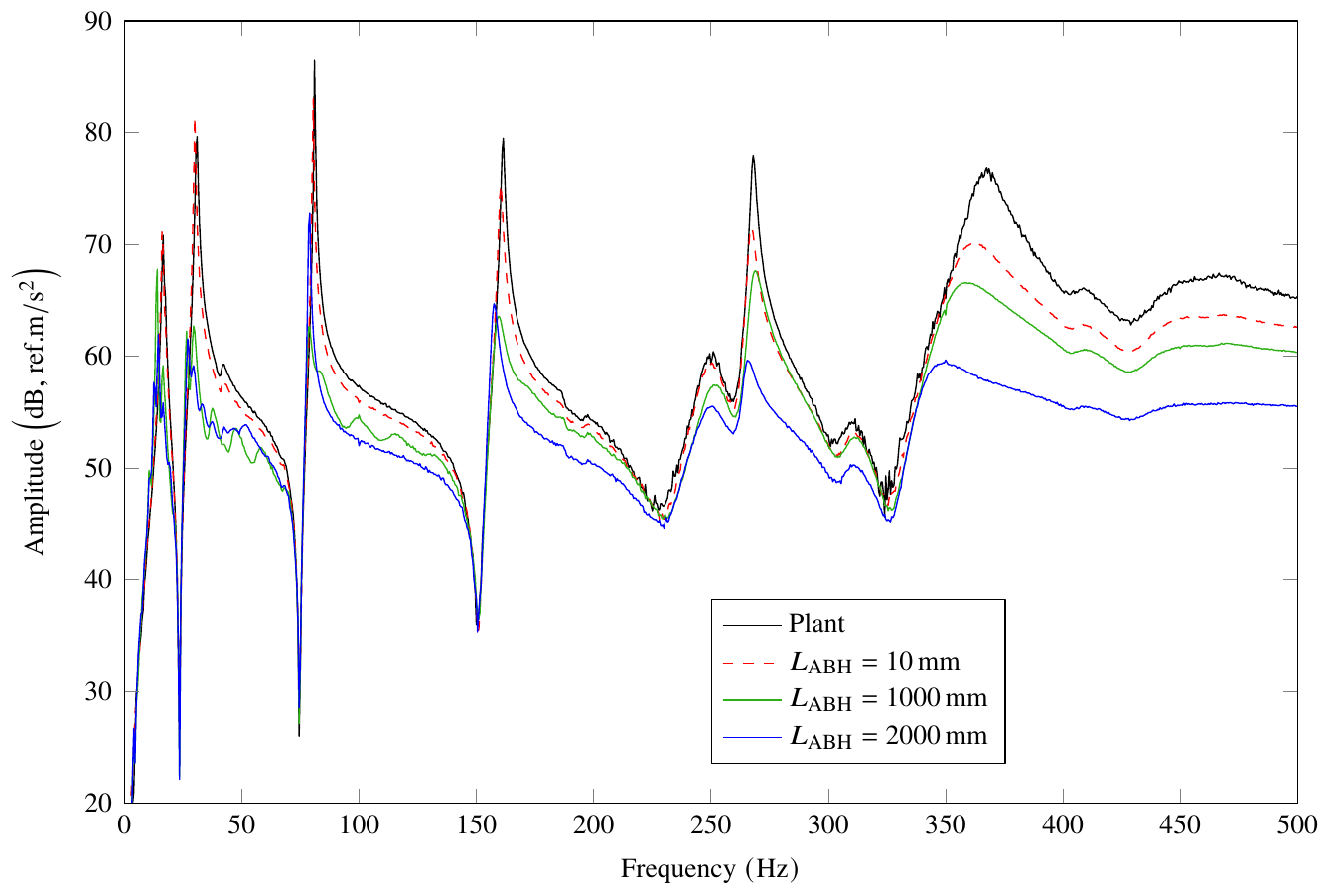}
	\caption{Influence of the ABH length on ABH performance.}
	\label{fig:influence length}
\end{figure}

Next, the length $L_{\mathrm{ABH}}$ is set to either $\SI{10}{\milli\meter}$, $\SI{1000}{\milli\meter}$, or $\SI{2000}{\milli\meter}$ in Figure~\ref{fig:influence length}. Modifying the length changes the cut-on frequency $f_{\mathrm{cut}}$ above which the ABH becomes effective. It is equal to $\SI{86}{\kilo\hertz}$, $\SI{8.6}{\hertz}$ and $\SI{2.15}{\hertz}$, respectively. For $\SI{10}{\milli\meter}$, the response amplitude is somewhat greater than the plant amplitude at low frequencies. This feature was already observed in Figure~\ref{fig:damping caract}; the ABH system had a lower modal damping than the cantilever beam for the first mode. Beyond $\SI{350}{\hertz}$, vibration reduction is achieved but should rather be attributed to the low-pass filter and the Craig-Bampton reduction. For the longest ABH beam, the resonance amplitude of all modes is further reduced compared to the nominal case. This is especially true for the first mode whose attenuation amounts to $\SI{9}{\decibel}$. A $\SI{3000}{\milli\meter}$-long VABH was also tested successfully but did not exhibit better performance than the $\SI{2000}{\milli\meter}$-long ABH.

%This last observation comes from the Craig-Bampton reduction. As observed in Figure~\ref{subfig: CB expe}, when fewer modes of Craig-Bampton are kept in the basis, the performance at high frequencies is improved. In our current situation the same reduction is applied on all cases, and the case $L_{\mathrm{ABH}}=\SI{2000}{\milli\meter}$ has lower resonant frequencies than the $L_{\mathrm{ABH}}=\SI{1000}{\milli\meter}$ system (the longer the beam, the lower the resonant frequencies). As a consequence, the last system better captures the ABH effect at high frequencies and presents a higher amplitude as in Figure~\ref{subfig: CB expe}.

\begin{figure}[H]
	\centering
	\includegraphics[scale=1]{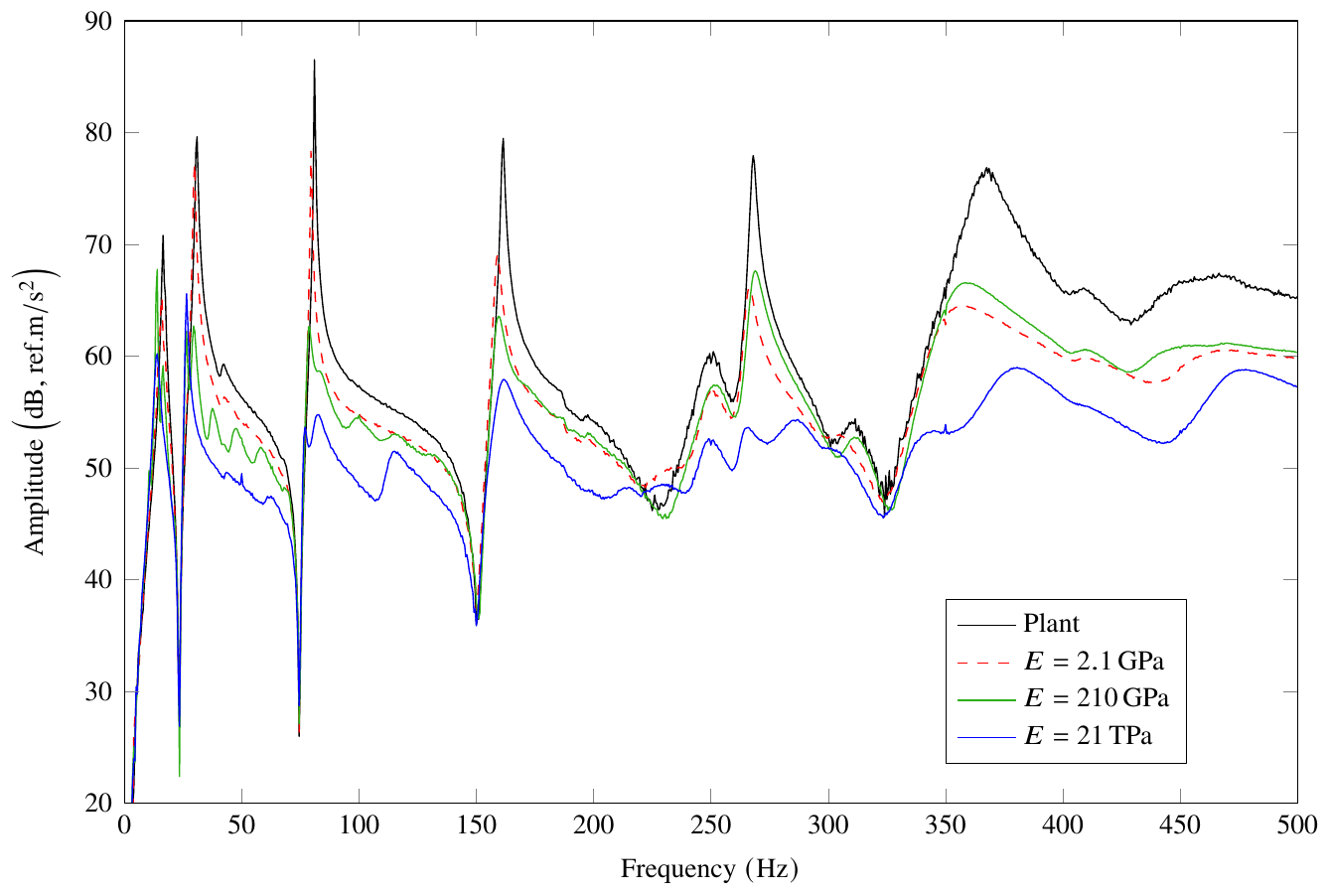}
	\caption{Influence of Young's modulus on ABH performance.}
	\label{fig:influence Young}
\end{figure}

Finally, three values of Young's modulus of the tapered wedge beam $E_{\mathrm{ABH}}$ are considered in Figure~\ref{fig:influence Young}, namely $\SI{2.1}{\giga\pascal}$, $\SI{210}{\giga\pascal}$, and $\SI{21}{\tera\pascal}$. For the greatest value, extremely good vibration attenuation is observed for all modes. In this case, the first mode is attenuated by $\SI{11}{\decibel}$.

\section{Conclusion}

This paper proposes the new concept of a VABH  which replaces a physical tapered wedge beam (i.e., an ABH) with a digital controller. The controller is designed so as to reproduce the ABH's dynamics at the interface with the host structure. Due to practical limitations, the VABH proposed herein only synthesizes the ABH's transversal dynamics. However, we demonstrated both numerically and experimentally that the resulting VABH exhibits performance and features which are very similar to those of a mechanical ABH. Specifically, our experiments highlighted that all the considered resonance peaks of the cantilever beam including the one below $\SI{20}{\hertz}$ were attenuated by at least $\SI{11}{\decibel}$.

Virtualizing the ABH solves important practical issues related to size, fatigue and manufacturing. Moreover, ABHs which could not even be imagined in practice, e.g., ABHs with very high Young's modulus or damping ratios, could be tested during the experimental campaign. The VABH has also its own limitations. First, stability margins must be verified to ensure the stability and robustness of the system. Second, the real-time controller limits the number of poles and zeros of the VABH, forcing us to adopt a Craig-Bampton reduction approach.

Future research could investigate VABHs applied to more complex host structures, implemented through piezoelectric transducers or VABHs synthesizing the rotational dynamics.

\section*{Acknowledgement}
This research is supported by a grant from the Belgian National Science Foundation (FRS-FNRS PDR T.0124.21), which is gratefully acknowledged.

\bibliography{biblio}

\end{document}